\documentclass[a4paper,conference]{IEEEtran}
\usepackage{times}
\usepackage{epsfig}
\usepackage{graphicx}
\usepackage{amsmath}
\usepackage{amssymb}

\usepackage{url}            % simple URL typesetting
\usepackage{booktabs}       % professional-quality tables
\usepackage{amsfonts}       % blackboard math symbols
\usepackage{nicefrac}       % compact symbols for 1/2, etc.
\usepackage{color}
\usepackage{amsbsy}

\begin{document}

\title{Boundary Defense Against Black-box Adversarial Attacks}
\author{\IEEEauthorblockN{Manjushree B. Aithal and Xiaohua Li}
\IEEEauthorblockA{Department of Electrical and Computer Engineering\\
Binghamton University\\
Binghamton, NY 13902, USA\\
Email: \{maithal1, xli\}@binghamton.edu}}

\maketitle

\begin{abstract}

Black-box adversarial attacks generate adversarial samples via iterative optimizations using repeated queries. Defending deep neural networks against such attacks has been challenging.  In this paper, we propose an efficient Boundary Defense (BD) method which mitigates black-box attacks by exploiting the fact that the adversarial optimizations often need samples on the classification boundary. Our method detects the boundary samples as those with low classification confidence and adds white Gaussian noise to their logits. The method's impact on the deep network's classification accuracy is analyzed theoretically. Extensive experiments are conducted and the results show that the BD method can reliably 
defend against both soft and hard label black-box attacks. It outperforms a list of existing defense methods.
For IMAGENET models, by adding zero-mean white Gaussian noise with standard deviation 0.1 to logits when the classification confidence is less than 0.3, the defense reduces the attack success rate to almost 0 while limiting the classification accuracy degradation to around 1 percent.

%Based on our findings, we selected the threshold up to 0.3 for ImageNET-based models and up to 0.7 for MNIST and CIFAR-based models. In the second step, the BD was deployed on various attack methods (both soft and hard label attacks) to evaluate the performance of the defense. The experimental analysis revealed that the addition of Gaussian noise reduced the attack success rate. It was also observed that increasing the magnitude of random noise immensely reduces the attacker's efficiency. In the last stage, we conducted a comparative study with existing defense methods and adaptive attack methods, to verify the effectiveness of BD against a list of state-of-the-art defense and black-box attack methods.

\end{abstract}

% \fi
\IEEEpeerreviewmaketitle

\section{Introduction }

Deep neural networks (DNNs) have achieved increasing demand in many practical applications \cite{chen2015deepdriving}\cite{hinton2012deep}\cite{krizhevsky2012imagenet}. However, studies over the past few years have also shown intriguing issue that DNN models are very sensitive and vulnerable to adversarial samples \cite{szegedy2013intriguing}\cite{biggio2013evasion}, implying potential security threats to their applications.

One of the widely studied adversarial attacks is the evasion attack, where the main aim of the attacker is to cause misclassification in the DNN model. 
%These goals can be met by using either untargeted or targeted attack approaches. In untargeted attack case, the attacker aims to cause any form of misclassification, whereas for the targeted attack method, the attacker manipulates the output of the DNN to a specific target label. 
%This attack has been widely studied in the setting of both untargeted attack and targeted attacks.
Black-box evasion attacks have attracted increasing research interests recently, where black-box means that the attacker does not know the DNN model but can query the model to get the DNN inference outputs, either the detailed confidence score or just a classification label \cite{alzantot2019genattack}\cite{brendel2017decision}\cite{chen2017zoo}\cite{tu2019autozoom}\cite{ilyas2018black}\cite{cheng2019improving}\cite{cheng2018query}\cite{cheng2019sign}\cite{li2019nattack}\cite{chen2020hopskipjumpattack}\cite{guo2019simple}. If the attacker has access to the full output logit values, they can apply soft-label attack algorithms such as \cite{chen2017zoo}\cite{tu2019autozoom}\cite{ilyas2018black}\cite{alzantot2019genattack}\cite{guo2019simple}. On the other hand, if the attacker has access to only the classification label, they can apply hard-label attack algorithms such as \cite{brendel2017decision}\cite{cheng2019sign}\cite{chen2020hopskipjumpattack}. 

Along with the surge of attack algorithms, there has been an increase in the development of defense algorithms such as Adversarial Training (AT) \cite{tramer2017ensemble}, input transformation \cite{buckman2018thermometer}\cite{samangouei2018defense}, gradient obfuscation \cite{papernot2016distillation}, and stochastic defense via randomization \cite{he2019parametric}\cite{wang2019protecting}\cite{qin2021random}\cite{nesti2021detecting}\cite{liang2018detecting}\cite{fan2019integration}\cite{li2018certified}. However, limitations of existing defense techniques have also been observed \cite{athalye2018obfuscated}\cite{carlini2017adversarial}\cite{carlini2017towards}. %\cite{geirhos2018imagenet}\cite{sokolic2017generalization}. 
It has been proven that stochastic defense suffers from large degradation of DNN performance or limited defense performance.
Gradient obfuscation method has also been proven to be ineffective. 

In this work, we develop an efficient and more effective method to defend the DNN against black-box attacks. 
%The primary goal of these attack methods is to reduce the accuracy of the model by manipulating the confidence score of the original input image. 
During the adversarial attack's optimization process, there is a stage that the adversarial samples are on the DNN's classification boundary. Boundary Defense BD$(\theta, \sigma)$, our method, detects these boundary samples as those with the classification confidence score below the threshold $\theta$ and adds white Gaussian noise with standard deviation $\sigma$ to their logits. This will prevent the attackers from optimizing their adversarial samples and maintain low DNN performance degradation. 

Major contributions of this work are:
\begin{itemize}
\item A new boundary defense algorithm BD$(\theta, \sigma)$ is developed, which can be implemented efficiently and mitigate reliably both soft and hard label black-box attacks.
\item Theoretical analysis is conducted to study the impact of the parameters $\theta$ and $\sigma$ on the classification accuracy.
\item Extensive experiments are conducted, which demonstrate that BD(0.3, 0.1) (or BD(0.7, 0.1)) reduces attack success rate to almost 0 with around 1\% (or negligible) classification accuracy degradation over the IMAGENET (or MNIST/CIFAR10) models. The defense performance is shown superior over a list of existing defense algorithms.
\end{itemize}

The organization of this paper is as follows. In Section \ref{relatedwork}, related works are introduced. In Section \ref{analysis}, the BD method is explained. In Section \ref{experiment}, experiment results are presented. Finally, conclusions are given in Section \ref{conclusion}.

%We investigated different threshold values to evaluate its effect on the model accuracy. For further study, a range of threshold values were shortlisted, which had minimum degradation in the model's accuracy. We performed extensive experimental analysis to understand the robustness of our defense method against state-of-the-art black-box attacks. To evaluate the robustness of the BD method against different strategies of an attacker we also evaluated the defense performance against the adaptive attack method (EOT) \cite{athalye2018obfuscated}. After our thorough analysis, to the best of our knowledge, the BD method has not been studied and reported. 

%%%%%%%%%%%%%%%%%%%%%%%%%%%%%%%%%%%%%%%%%%%%%%%%%%%%%%%%%%%%%%%%%%%%%%%%%%%%%%%%%%
\section{Related Work} \label{relatedwork}

%White-box attacks use the knowledge of DNN to calculate gradient and use the gradient to optimize adversarial samples. The methods include FGSM \cite{goodfellow2014explaining}, BIM \cite{kurakin2016adversarial}, PGD \cite{madry2017towards}, DeepFool \cite{moosavi2016deepfool}, C\&W attack \cite{carlini2017adversarial}, etc. Strong white-box attack algorithms are often used to evaluate the robustness to defense algorithms, such as the PGD and the more recently proposed auto-attack algorithm \cite{croce2020reliable}. 

%White-box defenses in general try to prevent the successful gradient calculation of the attacker by transforming the input (such as JPEG compression \cite{dziugaite2016study}), randomizing the input or the network \cite{xie2018mitigating}. However, such gradient obfuscation based methods have been shown in general not robust \cite{carlini2017towards}\cite{athalye2018obfuscated}\cite{tramer2020adaptive}. In contrast, adversarial training \cite{szegedy2013intriguing}\cite{madry2017towards}\cite{wong2020fast} is more robust but unfortunately limited by computational complexity or unforeseen attack methods. 

Black-box adversarial attacks can be classified into soft-label and hard-label attacks. In soft-label attacks like AutoZOOM \cite{chen2017zoo}\cite{cheng2018query} and NES-QL \cite{ilyas2018black}, the attacker generates adversarial samples using the gradients estimated from queried DNN outputs. In contrast, SimBA \cite{guo2019simple}, GenAttack \cite{alzantot2019genattack} and Square Attack \cite{andriushchenko2020square} resort to direct random search to obtain the adversarial sample.
Hard-label attacks like NES-HL \cite{ilyas2018black}, BA (Boundary Attack) \cite{brendel2017decision}, Sign-OPT \cite{cheng2018query}\cite{cheng2019sign}, and HopSkipJump \cite{chen2020hopskipjumpattack} start from an initial adversarial sample and iteratively reduce the distance between the adversarial sample and original sample based on the query results.

For the defense against black-box attacks, a lot of methods are derived directly from the defense methods against white-box attacks, such as input transformation \cite{dziugaite2016study}, network randomization \cite{xie2018mitigating} and adversarial training \cite{tramer2020adaptive}. The defenses designed specifically for black-box attack, are denoised smoothing \cite{salman2020denoised}, malicious query detection \cite{chen2020stateful}\cite{li2020blacklight}\cite{pang2020advmind}, and random-smoothing \cite{cohen2019certified}\cite{salman2019provably}. Nevertheless, their defense performance is not reliable and defense cost or complexity is too high. %Along with these methods, there is a  focus on the idea of training robust models with reasonable computational budget \cite{croce2021robustbench}. 
Adding random noise to defend against black-box attacks has been studied recently as a low-cost approach, where \cite{byun2021small}\cite{qin2021theoretical}\cite{xie2018mitigating} add noise to the input, and \cite{lecuyer2019certified} \cite{liu2018towards}\cite{he2019parametric} add noise to input or weight of each layer. Unfortunately, heavy noise is needed to defend against hard-label attacks (in order to change hard labels) but heavy noise leads to severe degradation of DNN accuracy. Our proposed BD method follows similar approach, but we add noise only to the DNN outputs of the boundary samples, which makes it possible to apply heavy noise without significant degradation in DNN accuracy. 

%Other techniques such as PixelDP \cite{lecuyer2019certified} and random-smoothing \cite{cohen2019certified}\cite{salman2019provably} were proposed to train the model using Gaussian noise. But this introduces a common trade-off between noise level, model accuracy and also the requirement of a large number of predictions. Along with these methods, majority of research community focuses on the ideas of training robust models with reasonable computational budget \cite{croce2021robustbench} (i.e., included in the RobustBench library). The robustness of the pre-trained model listed in RobustBench library are evaluated with an ensemble of white-box and black-box attacks. Considering all these issues and setback from the previous works, we have made sure that the model degradation doesn't occur in our proposed BD method. 

%%%%%%%%%%%%%%%%%%%%%%%%%%%%%%%%%%%%%%%%%%%%%%%%%%%%%%%%%%%%%%%%%%%%%%%%%%%%%%%%%%%%%%%%%%%%%%%%%%%%%%%%%%%%%%%%
\section{Boundary Defense} \label{analysis}

\subsection{Black-box attack model}
Consider a DNN that classifies an image ${\bf X}_0$ into class label $c$ within $N$ classes. The DNN output is softmax logit (or confidence score) tensor $F({\bf X}_0)$. The classification result is $c=\arg\max_{i} {F}_i({\bf X}_0)$, where $F_i$ denotes the $i$th element function of $F$, $i=0, \cdots, N-1$. The attacker does not know the DNN model but can send samples ${\bf X}$ to query the DNN and get either $F({\bf X})$ or just $c$. The objective of the attacker is to generate an adversarial sample ${\bf X}={\bf X}_0+\Delta {\bf x}$ such that the output of the classifier is $t = \arg\max_{i} F_i({\bf X}) \neq c$, where the adversary $\Delta {\bf x}$ should be as small as possible. 

\textbf{Soft-Label Black-box Attack:} The attacker queries the DNN to obtain the softmax logit output tensor ${F}({\bf X})$. With this information, the attacker minimizes the loss function $f_{{\rm SL}}({\bf X})$ for generating the adversarial sample ${\bf X}$ \cite{tu2019autozoom},
\begin{equation}
    f_{{\rm SL}}({\bf X}) = {\cal D}({\bf X}, {\bf X}_0) + \lambda {\cal L}(F({\bf X}), t),   \label{eq1.10}
\end{equation}
where ${\cal D}(\cdot, \cdot)$ is a distance function, e.g., $\| {\bf X}-{\bf X}_0 \|_p$, and ${\cal L}(\cdot, t)$ is the loss function, e.g., cross-entropy \cite{ilyas2018black} and C\&W loss \cite{carlini2017adversarial}. 
%The targeted attack (towards class $t$) in a black-box settings is achieved by minimizing the loss function
%\begin{equation}
%    f({\bf X})= \| {\bf X}-{\bf X}_0 \|^2 + \lambda \log \frac{F_{\max}({\bf X})}{F_t({\bf X})},  \label{eq2}
%\end{equation}
%where $F_{\max}({\bf X}) = \{F_i({\bf X}): i=\arg\max_j F_j({\bf X}), \forall j \neq t\}$ is the logit value of the largest non-target element, $F_t({\bf X})$ is the logit value of the target element, and $\lambda$ is the weighting hyper-parameter.
%These attack algorithm runs in two stages, the first stage is the adversarial search stage and the second stage is the adversarial fine-tuning stage. For our BD method's analysis, we focus on the first stage where the attacker is generating the adversarial sample ${\bf X}$ with classification result $F_{t}({\bf X})$, i.e., the targeting label. During this stage the main goal of the attack algorithm is to iteratively reduce the non-target element logit value below the target class logits.

\textbf{Hard-Label Black-box Attack}: The attacker does not use $F({\bf X})$ but instead uses the class label $\arg \max_i F_i({\bf X})$ to optimize the adversarial sample ${\bf X}$. A common approach for the attacker is to first find an initial sample ${\bf X}_{t,0}$ in the class $t \neq c$, i.e., $\arg\max_i F_i({\bf X}_{t,0})=t$. Then, starting from ${\bf X}_{t,0}$, the attacker iteratively estimates new adversarial samples ${\bf X}$ in the class  $t$ so as to minimize the loss function $f_{{\rm HL}}({\bf X}) = {\cal D}({\bf X}, {\bf X}_0)$.
%under the constraint $\lambda=0$ and $\arg\max_i F_i({\bf x}) = t$. 

The above model is valid for both targeted and untargeted attacks. The attacker's objective is to increase attack success rate (ASR), reduce query counts (QC), and reduce sample distortion ${\cal D}({\bf X}, {\bf X}_0)$. In this paper, we assume that the attacker has a large enough QC budget  and can adopt either soft-label or hard-label black-box attack algorithms. Thus, our proposed defense's main objective is to reduce the ASR to 0.

%\subsection{Principles of boundary defense}
%Note that the $\log$ term in (\ref{eq2}) is just the Carlini-Wagner white-box attack loss function $\max\{0, \max_{j\neq t} \log F_j({\bf x}) - \log F_t({\bf x}) \}$ used in \cite{tu2019autozoom}. The first $\max$ is skipped because we consider the adversarial search stage when the target has not been reached yet.

%From \cite{tu2019autozoom}, the attacker queries the 
%DNN with inputs ${\bf x}$ and ${\bf x}+ \beta {\bf u}_j$ where $\beta$ is a small constant and ${\bf u}_j$ is a direction tensor. The gradient estimated in adversarial search is 
%\begin{align}
%    {\bf g}_j =& \frac{1}{\beta}{\bf u}_j \left[ \|{\bf x} 
%     +\beta{\bf u}_j-{\bf x}_0\|^2 - \| {\bf x}-{\bf x}_0 \|^2 \right. \nonumber \\
%    &\left. + \lambda \log \frac{F_{\max}({\bf x}+\beta {\bf u}_j)/F_t({\bf x}+\beta {\bf u}_j)} {F_{\max}({\bf x})/F_t({\bf x})} \right].   \label{eq3}
%\end{align}
%The gradient estimation is ${\bf g}_j = a {\bf u}_j$, which is different from the true gradient what would be calculated with white-box attacks such as PGD. We can derive the SNR as follows.
%%%%%%%%%%%%%%%%%%%%%%%%%%%%%%%%%%%%%%%%%%%%%%%%%%%%%%%%%%%%%%

\begin{figure}[t]
	\centering
    \includegraphics[width=1\linewidth]{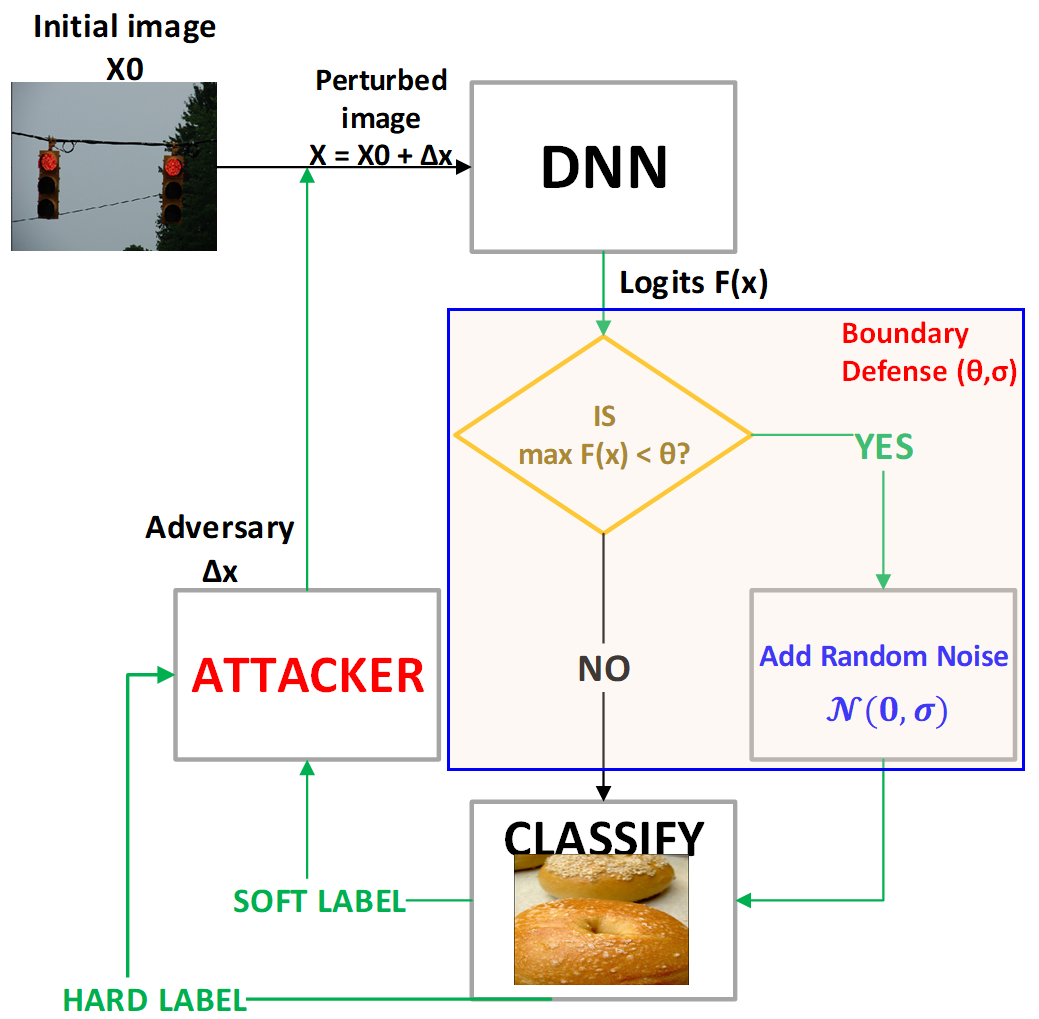}
	\caption{Schematic representation of the black-box attack and the Boundary Defense BD$(\theta, \sigma)$ (highlighted region).}
	\label{fig:model}
\end{figure}

%%%%%%%%%%%%%%%%%%%%%%%%%%%%%%%%%%%%%%%%%%%%%%%%%%%%%%%%%%%%%%%%%%%%%%%5
\subsection{Boundary Defense Algorithm}
In this work, we propose a Boundary Defense method that defends the DNN against black-box (both soft and hard label, both targeted and untargeted) attacks by preventing the attacker's optimization of $f_{{\rm SL}}({\bf X})$ or $f_{{\rm HL}}({\bf X})$. As illustrated in Fig. \ref{fig:model}, for each query of ${\bf X}$, once the defender finds that the classification confidence $\max F({\bf X})$ is less than certain threshold $\theta$, the defender adds zero-mean white Gaussian noise ${\cal N}(0, \sigma^2)$ with a certain standard deviation $\sigma$ to all the elements of $F({\bf X})$. The DNN softmax logits thus become
\begin{equation}
    F_{{\rm BD}}({\bf X}) = \left\{ \begin{array}{ll} F({\bf X}), & {\rm if} \; \max F({\bf X}) > \theta \\
    F({\bf X}) + V, & {\rm otherwise} \end{array} \right.    \label{eq2.1}
\end{equation}
% where $V \sim {\cal N}(0, \sigma^2 {\bf I})$ is a $C$ dimensional Gaussian vector with zero-mean and standard deviation $\sigma$. 
% can be calculated as
% \begin{equation}
%     P = \frac{F_{\max} ({\bf X})} {\sum_{i=1}^{C} F_i({\bf X})}
% \end{equation}
% if top-1 ACC is used, and
% \begin{equation}
%     P = \frac{\sum_{j=1}^K F_j ({\bf X})} {\sum_{i=1, i\neq j}^{C} F_i({\bf X})}
% \end{equation}, 
% if top-K ACC is used.
where $V \sim {\cal N}({\bf 0}, \sigma^2{\bf I})$ and ${\bf I}$ is an identity matrix. The DNN outputs softmax logits clip\{$F_{{\rm BD}} ({\bf X})$, 0, 1\} when outputting soft labels or its classification label $\arg \max_i F_{{\rm BD}, i}({\bf X})$ when outputting hard labels. 
% $F_{\max}({\bf X})$ denotes the logit (or softmax) output entry) that has the maximum value, whereas $F_j({\bf X})$ denote the top K  maximum elements. 

We call it the BD$(\theta, \sigma)$ algorithm because samples with low confidence scores are usually on the classification boundary. %the defender detects the boundary samples by comparing their classification confidence score $\max F({\bf X})$ with the pre-set threshold $\theta$ and adds white Gaussian noise ${\cal N}(0, \sigma^2)$ to the query outputs of the boundary samples. %Boundary samples can be easily detected because their maximum softmax logit values are usually small. 
For a well-designed DNN, the clean (non-adversarial) samples can usually be classified accurately with high confidence scores. Those with low confidence scores happen rarely and have low classification accuracy. In contrast, when the attacker optimizes $f_{{\rm SL}}({\bf X})$ or $f_{{\rm HL}}({\bf X})$, there is always a stage that the adversarial samples ${\bf X}$ have small $\max F({\bf X})$ values. 

For example, in the soft-label black-box targeted attacks, the attacker needs to maximize the $t$th logit value $F_t({\bf X})$ by minimizing the cross-entropy loss ${\cal L}(F({\bf X}), t)= -\log F_t({\bf X})$. Initially $F_t({\bf X}_0)$ is very small and $F_c({\bf X}_0)$ is large. The optimization increases $F_t({\bf X})$ while reducing $F_c({\bf X})$. There is a stage that all logit values are small, which means ${\bf X}$ is lying on the classification boundary.

As another example, a typical hard-label black-box targeted attack algorithm first finds an initial sample inside the target class $t$, which we denote as ${\bf X}_{t, 0}$. The algorithm often uses line search to find a boundary sample ${\bf X} = \alpha {\bf X}_{t, 0} + (1-\alpha) {\bf X}_0$ that maintains label $t$, where $\alpha$ is the optimization parameter. Then the algorithm randomly perturbs ${\bf X}$, queries the DNN, and uses the query results to find the direction to optimize $f_{{\rm HL}}({\bf X})$. Obviously, ${\bf X}$ must be on the decision boundary so that the randomly perturbed ${\bf X}$ will lead to changing DNN hard-label outputs. Otherwise, all the query results will lead to a constant output $t$, which is useless to the attacker's optimization process. 

Therefore, for soft-label attacks there is an unavoidable stage of having boundary samples and for hard-label attacks the boundary samples are essential. Our BD method exploits this weakness of black-box attacks by detecting these samples and scrambling their query results to prevent the attacker from optimizing its objective. 

%We coined this method as \textbf{\underline{``Boundary Defense"}} (BD) because as shown above, at a certain stage in optimization process, both the true label and target label logit values (and in fact all the logits) are small. This stage occurs at the decision boundary of the original sample and the target sample, where our proposed defense makes the attacker impossible to cross this boundary successfully. 

One of the advantages of the BD$(\theta, \sigma)$ algorithm is that it can be implemented efficiently and inserted into DNN models conveniently with minimal coding. Another advantage is that the two parameters $(\theta, \sigma)$ make it flexible to adjust the BD method to work reliably. Large $\theta$ and $\sigma$ lead to small ASR but significant DNN performance degradation. Some attacks are immune to small noise (small $\sigma$), such as the HopSkipJump hard-label attack \cite{chen2020hopskipjumpattack}. Some other attacks such as SimBA \cite{guo2019simple} are surprisingly immune to large noise in boundary samples, which means that simply removing boundary samples or adding extra large noise to boundary samples as suggested in \cite{chen2020hopskipjumpattack} does not work. The flexibility of $(\theta, \sigma)$ makes it possible for the BD method to deal with such complicated issues and to be superior over other defense methods.

%it can limit the DNN's classification performance degradation by appropriate $(\theta, \sigma)$. This makes our method superior over existing defense methods since defending against hard-label attacks need large noise or randomization (in order to change hard label outputs), which will lead to large DNN performance degradation if applied to all the samples. Finally

%Please note that, for a well trained DNN, its output over the clean input samples usually have high confidence-score $\max F_i({\bf X})$. The samples with low confidence-score are not reliable to classify. Adding noise to these samples will not degrade the performance of DNN too much if $\theta$ is chosen appropriately. As the third advantage, the BD method allows adding large noise with high $\sigma$, which is more effective to reduce the attacker's ASR, especially to hard-label black-box attacks. 
%The boundary images that the attacker generates during its optimization procedure are usually a mixture of clean and heavily distorted images (that are very different from the clean input images). The BD method can also be used in combination with some other existing defenders for a robust performance against the attacker's strategies and we have left this for the future study.
%As a comparison, some defense methods that rely on another deep network to classify the input images to be clean image or adversarial image can also be used for assistance. However, such extra deep network can be easily attacked by the attacker if the attacker knows this defense strategy.
%%%%%%%%%%%%%%%%%%%%%%%%%%%%%%%%%%%%%%%%%%%%%%%%%%%%%%

\subsection{Properties of Boundary Samples} \label{acc_degradation}
In this section, we study BD$(\theta, \sigma)$'s impact on the DNN's classification accuracy (ACC) when there is no attack, which provides useful guidance to the selection of $\theta$ and $\sigma$.

\begin{figure}[t]
	\centering
    \includegraphics[width=0.48\linewidth]{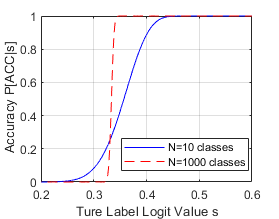}
    \includegraphics[width=0.48\linewidth]{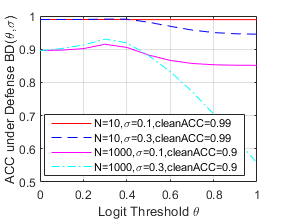}
    \centerline{(a) \hspace{0.4\linewidth} (b)}
	\caption{Impact of parameters $\theta$ and $\sigma$ to classification accuracy (ACC). (a) ACC as function of true logit value $s = F_c({\bf X})$. (b) ACC when boundary defense BD$(\theta, \sigma)$ is applied. CleanACC is the DNN's ACC without attack/defense.}
	\label{fig:accanalysis}
\end{figure}

Consider a clean sample ${\bf X}$ with true label $c$ and confidence $s = F_c({\bf X})$. Since the DNN is trained with the objective of maximizing $F_c({\bf X})$, we can assume that all the other logit values $F_i({\bf X})$, $i=0, \cdots, N-1$, $i \neq c$, are independent and identically distributed uniform random variables with values within $0$ to $a=(1-s)/(N-1)$, i.e., $F_i({\bf X}) \sim U(0, a)$. Without loss of generality, let $F_0({\bf X})$ be the maximum among these $N-1$ values. Then $Y = \sum_{i=1, i\neq c}^{N-1} F_i({\bf X})$ follows Irwin-Hall distribution with cumulative distribution function (CDF)
\begin{equation}
    P_Y[y < x] = \frac{1}{(N-2)!} \sum_{k=0}^{\lfloor x/a \rfloor} (-1)^k \left(\begin{array}{c} N-2 \\ k \end{array} \right) \left( \frac{x}{a} - k \right)^{N-2}.   \label{eq3.9}
\end{equation}
When $N$ is large, the distribution of $Y$ can be approximated as normal ${\cal N}\left( \frac{(1-s)(N-2)}{2(N-1)}, \left( \frac{1-s}{N-1} \right)^2 \frac{N-2}{12} \right)$. We denote its CDF as $\Phi(x)$. Since the sample ${\bf X}$ is classified accurately if and only if $s > F_0({\bf X})$, the classification accuracy $P[{\rm ACC}|s]$ can be derived as
\begin{equation}
    P[{\rm ACC}|s] = P_Y[y > 1-2s] = 1 - \Phi(1-2s). \label{eq3.10}
\end{equation}
Using (\ref{eq3.10}), we can calculate $P[{\rm ACC}|s]$ for each $s$, as shown in Fig. \ref{fig:accanalysis}(a) for $N=10$ and $1000$ classes. It can be seen that for $N=1000$, if $s < 0.32$, then the sample's classification ACC is almost 0. This means that we can set $\theta \leq 0.32$ to safely scramble all those queries whose maximum logit value is less than 0.32 without noticeable ACC degradation. %For $N=10$, we need to use larger $\theta$ and together with a relatively small $\sigma$ to reduce ACC degradation. %samples with true logit values $s<0.4$ may not be classified reliably, so we can set the threshold $\theta$ to larger values. 

Next, to evaluate the ACC when BD$(\theta, \sigma)$ is applied, we assume the true label $c$'s logit value $s$ follow approximately half-normal distribution, whose probability density function is
\begin{equation}
    f_S(s) = \left\{ \begin{array}{ll}
    \frac{\sqrt{2}}{\nu \sqrt{\pi}} e^{-\frac{(1-s)^2}{2\nu^2}}, & s \leq 1 \\
    0, & {\rm otherwise} \end{array} \right. \label{eq3.14}
\end{equation}
with the parameter $\nu$.
The ACC of the DNN without attack or defense (which we call cleanACC) is then
\begin{equation}
    {\rm cleanACC} = \int_0^1 P[{\rm ACC}|s]f_S(s) ds.  \label{eq3.15}
\end{equation}
Using (\ref{eq3.14})-(\ref{eq3.15}), we can find the parameter $\nu$ for each clean ACC. For example, for $N=1000$ and a DNN with clean ACC 90\%, the distribution of true logit $s$ follows (\ref{eq3.14}) with $\nu=0.41$. 

When noise is added, each $F_i(X)$ becomes $F_i(X)+v_i$ for noise $v_i \sim {\cal N}(0, \sigma^2)$. Following similar derivation of (\ref{eq3.9})-(\ref{eq3.10}), we can obtain the ACC of the noise perturbed logit $F_c(X)+v_c$ as
\begin{equation}
    P[{\rm ACC}|s,\sigma] = 1 - \tilde{\Phi}(1-2s),
\end{equation}
where $\tilde{\Phi}$ is the CDF of the new normal distribution 
${\cal N}\left( \frac{(1-s)(N-2)}{2(N-1)}, \left( \frac{1-s}{N-1} \right)^2 \frac{N-2}{12} + (N+2)\sigma^2 \right)$. The ACC under the defense is then
\begin{equation}
    {\rm ACC} = \int_0^{\theta} P[{\rm ACC}|s,\sigma]f_S(s)ds + \int_{\theta}^1 P[{\rm ACC}|s] f_S(s) ds.
\end{equation}
Fig. \ref{fig:accanalysis}(b) shows how the defense ACC degrades with the increase of $\theta$ and $\sigma$. We can see that with $\sigma=0.1$, there is almost no ACC degradation for $N=10$. For $N=1000$, ACC degradation is very small when $\theta<0.4$ but grows to 5\% when $\theta>0.6$. Importantly, under $\theta<0.4$ we can apply larger noise $\sigma=0.3$ safely without obvious ACC degradation. This shows the importance of scrambling boundary samples only. Existing defenses scramble all the samples, which corresponds to $\theta=1$, and thus suffer from significant ACC degradation.

%%%%%%%%%%%%%%%%%%%%%%%%%%%%%%%%%%%%%%%%%%%%%%%%%%%%%%%%%%%%%%%%%%%%%%%%
\section{Experiments}
\label{experiment}
%Our experiments focused on following ideas: 1) determining best threshold $\theta$ and Gaussian random noise $\sigma$, that has less degradation in model accuracy;
%2) understanding the effectiveness of the BD method against existing state-of-the-art attack methods; 3) performance evaluation of the BD method in adaptive attacks setting; and 4) comparison of our defense with other existing adversarial defenses. 

\subsection{Experiment Setup}\label{setup}

In the first experiment, with the full validation datasets of MNIST (10,000 images), CIFAR10 (10,000 images), IMAGENET (50,000 images) 
we evaluated the degradation of classification accuracy of a list of popular DNN models when our proposed BD method is applied.

In the second experiment, with $1000$ validation images of MNIST/CIFAR10 and $100$ validation images of IMAGENET, we evaluated the defense performance of our BD method against several state-of-the-art black-box attack methods, including soft-label attacks {\bf AZ} (AutoZOOM) \cite{tu2019autozoom}, {\bf NES-QL} (query limited) \cite{ilyas2018black}, {\bf SimBA} (SimBA-DCT) \cite{guo2019simple}, and {\bf GA} (GenAttack) \cite{alzantot2019genattack}, as well as hard-label attacks {\bf NES-HL} (hard label) \cite{ilyas2018black}, {\bf BA} (Boundary Attack) \cite{brendel2017decision}, {\bf HSJA} (HopSkipJump Attack) \cite{chen2020hopskipjumpattack}, and {\bf Sign-OPT} \cite{cheng2019sign}. %Experiments were conducted over the MNIST, CIFAR10 and IMAGENET datasets. 
We adopted their original source codes with the default hyper-parameters and just inserted our BD$(\theta, \sigma)$ as a subroutine to process $F({\bf X})$ after each model prediction call. These algorithms used the InceptionV3 or ResNet50 IMAGENET models. To maintain uniformity and fair comparison, we considered the $l_{2}$ norm setting throughout the experiment.

We also compared our BD method with some representative black-box defense methods, including {\bf NP} (noise perturbation), {\bf JPEG} compression, {\bf Bit-Dept}, and {\bf TVM} (Total Variation Minimization), whose data were obtained from \cite{guo2017countering}, for soft-label attacks, and {\bf DD} (Defensive Distillation) \cite{papernot2016distillation}, {\bf Region-based} classification \cite{cao2017mitigating}, and {\bf AT} (Adversarial Training) \cite{goodfellow2014explaining} for hard-label attacks. 

In order to have a more persuasive and comprehensive study of the robustness of the proposed BD method, we also performed experiments using Robust Benchmark models \cite{croce2021robustbench}, such as {\bf RMC} (Runtime Masking and Cleaning) \cite{wu2020adversarial}, {\bf RATIO} (Robustness via Adversarial Training on In- and Out-distribution)\cite{augustin2020adversarial}, {\bf RO} (Robust Overfitting)\cite{rice2020overfitting}, {\bf MMA} (Max-Margin Adversarial)\cite{ding2018mma}, {\bf ER} (Engstrom Robustness)\cite{engstrom2019adversarial}, {\bf RD} (Rony Decoupling)\cite{rony2019decoupling}, and {\bf PD} (Proxy Distribution)\cite{sehwag2021improving} models, over the CIFAR10 dataset for various attack methods. 

As the primary performance metrics, we considered {\bf ACC} (DNN's classification accuracy) and {\bf ASR} (attacker's attack success rate). The ASR is defined as the ratio of samples with $\arg \max_i F_i({\bf X}) = t \neq c$. Without defense, the hard-label attack algorithms always output adversarial samples successfully with the label $t$ (which means ASR = 100\%). Under our defense the ASR will be reduced due to the added noise, so ASR is still a valid performance measure. On the other hand, since most hard-label attack/defense papers use the ASR defined as the ratio of samples satisfying both $\arg \max_i F_i({\bf X}) = t$ and median $l_2$ distortion ($\sqrt{\| {\bf X}-{\bf X}_0 \|^2/M}$ when ${\bf X}_0$ has $M$ elements) less than a certain threshold, 
we will also report our results over this ASR, which we called {\bf ASR2}. 
% \begin{equation}
%     l_2 = median(\sqrt{(X - X_0)^2/T})
%     \label{l2 eq}
% \end{equation}
% where, X is adversarial example, $X_0$ is original image, T is total number of test images. 

We show only the results of targeted attacks in this section. Experiments of untargeted attacks as well as extra experiment data and result discussions are provided in supplementary material.

\subsection{ACC Degradation Caused by Boundary Defense}\label{accevaluation}

\begin{figure}[t]
	\centering
    \includegraphics[width=0.9\linewidth]{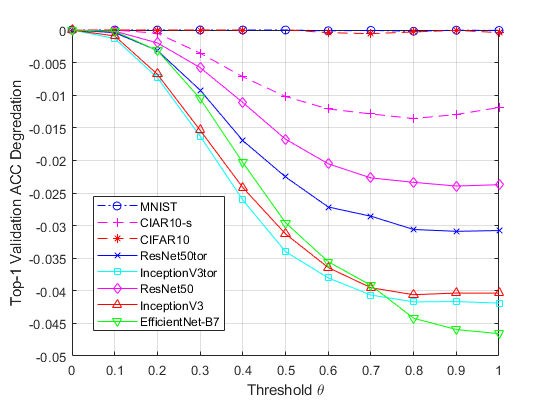}
    % \centerline{(a) Top-1 ACC degradation.}
    % \includegraphics[width=1\linewidth]{figures/ACC4fig1s01top5.png}
    % \centerline{(b) To-5 ACC degradation.}
	\caption{Top-1 classification accuracy degradation (Defense ACC $-$ Clean ACC) versus $\theta$. $\sigma=0.1$.} %(a) Top-1 ACC. (b) Top-5 ACC. }
	\label{fig:accdegradation}
\end{figure}

For MNIST, we trained a 5-layer convolutional neural network (CNN) with clean ACC 99\%. For CIFAR10, we trained a 6-layer CNN with clean ACC 83\% and also applied the pre-trained model of \cite{xu2019pc} with ACC 97\%, which are called {\bf CIFAR10-s} and {\bf CIFAR10}, respectively. For IMAGENET, we used standard pre-trained models from the official Tensoflow library ({\bf ResNet50}, {\bf InceptionV3}, {\bf EfficientNet-B7}) and the official PyTorch library ({\bf ResNet50tor}, {\bf InceptionV3tor}), where ``-tor" indicates their PyTorch source. 

We used the validation images to query the DNN models and applied our BD algorithm to modify the DNN outputs before evaluating classification ACC.
It can be observed from Fig. \ref{fig:accdegradation} that with $\theta \leq 0.3$ we can keep the loss of ACC around $1\%$ for IMAGENET models (from 0.5\% of ResNet50 to 1.5\% of InceptionV3). $\theta > 0.6$ leads to near 5\% ACC degradation. For MNIST and CIFAR10 the ACC has almost no degradation, but CIFAR10-s has limited 1.5\% ACC degradation for large $\theta$.
This fits well with the analysis results shown in Fig. \ref{fig:accanalysis}(b). Especially, most existing noise defense methods, which don't exploit boundary (equivalent to $\theta = 1$), would result in up to 5\% ACC degradation for IMAGENET models. 

\subsection{Performance of BD Defense against Attacks}

%We implemented our defense method against the black-box attack algorithms and considered the attack success rate (ASR) and the median distortion as the prime evaluation metric. After varying the boundary threshold ($\theta$), we also varied the standard deviation of the Gaussian random noise ($\sigma \in \{0.01, 0.05, 0.1\}$) that is used for selective randomization of queried DNN output after the attack threshold reached. 

\subsubsection{ASR of soft-label black-box attacks}\label{softlabel section}

\begin{figure}[ht]
	\centering
    \includegraphics[width=1\linewidth]{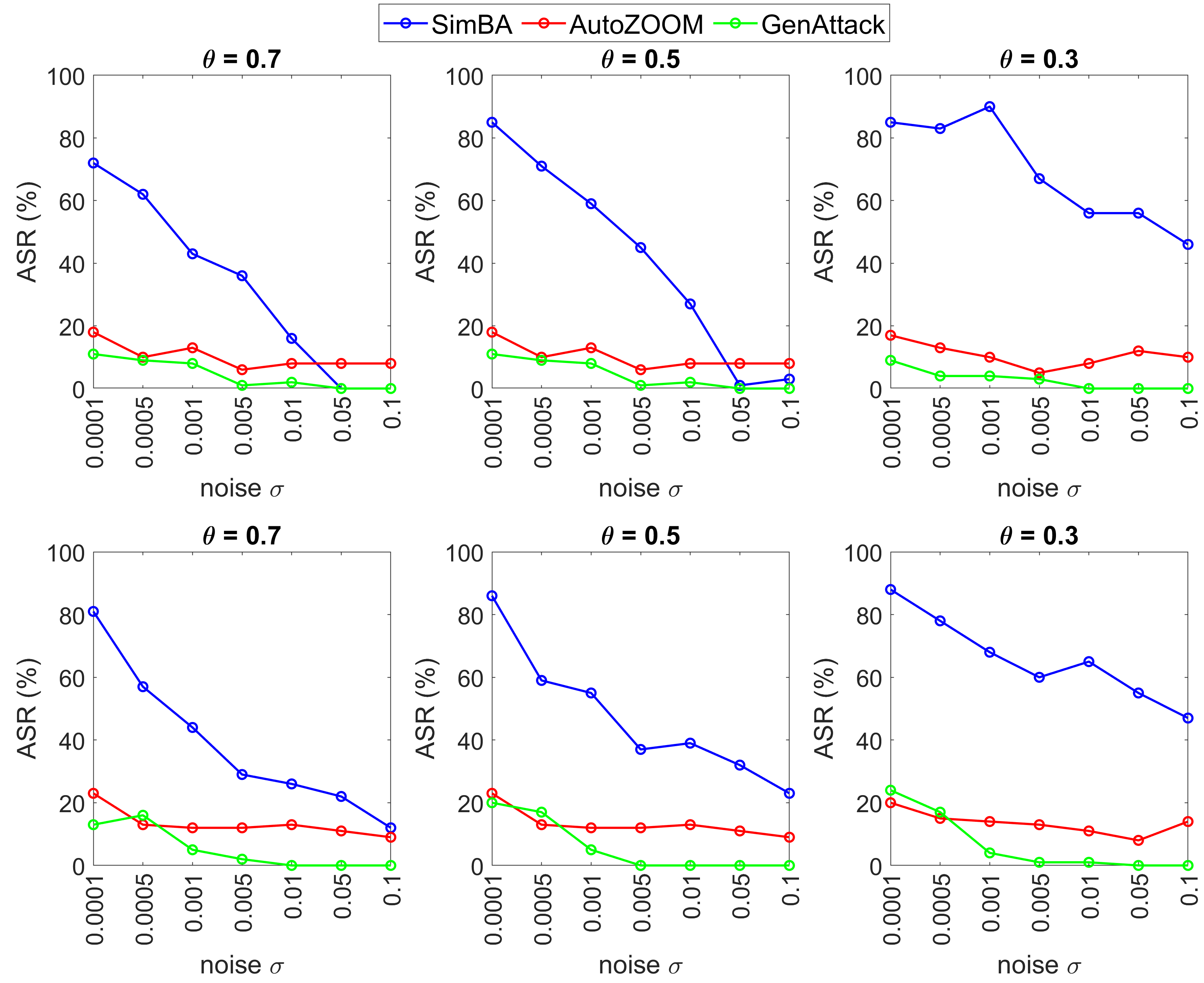}
	\caption{ASR(\%) vs noise level $\sigma$ for various boundary threshold $\theta$. The top row is for MNIST, and the bottom row is for CIFAR10.}
	\label{fig:soft-label asr}
\end{figure}

Table \ref{tbl:ASR1} shows the ASR of soft-label black-box attack algorithms under our proposed BD method. To save the space we have shown the data of $\sigma=0.1$ only. Results regarding varying $\theta$ and $\sigma$ are shown in Fig. \ref{fig:soft-label asr}. 
%Note that in the graphs we have considered the best $\theta$ values calculated from section \ref{accevaluation}.
\begin{table}[t]
	\caption{ASR (\%) of Targeted Soft-Label Attacks. $\sigma$ = 0.1.}
	\label{tbl:ASR1}
	\centering
	\begin{tabular}{c|c|ccc}
	
		\toprule
		Dataset     & Attacks   %&  & ASR (\%) &\\ & 
		& No defense & $\theta_1$ = 0.5 & $\theta_2$ = 0.7 \\
		      % \cmidrule{3-5}
		     % &  &  &  & \\
		\midrule
		 & AZ & 100  & 8 & 8  \\
		  MNIST    & GA & 100 & 0 & 0\\
		       & SimBA & 97 & 3 & 0  \\
		 \midrule
		     &  AZ  & 100 & 9 & 9  \\
		  CIFAR10    & GA & 98.76 & 0 & 0   \\
		%  & NES & 100 &  &  \\
		      & SimBA & 97.14 & 23 & 15\\
		 \midrule \midrule
		 Dataset     & Attacks   %&  &ASR (\%) &\\
		     %&   
		     & No Defense & $\theta_1$ = 0.1 & $\theta_2$ = 0.3 \\
		  \midrule
		  & AZ & 100  & 0 & 0 \\
		  IMAGENET    & NES-QL & 100  & 69 & 8 \\
		      & GA & 100  & 0 & 0\\
		      & SimBA & 96.5  & 6 & 2 \\
		\bottomrule
	\end{tabular}
\end{table}

From Table \ref{tbl:ASR1} we can see that with the increase in $\theta$, the ASR of all the attack algorithms drastically reduced. 
Over the IMAGENET dataset, the BD method reduced the ASR of all the attack algorithms to almost $0$ with $(\theta, \sigma) = (0.3, 0.1)$. For MNIST/CIFAR10 datasets, the BD method with $(\theta, \sigma)=(0.5, 0.1)$ was enough. Fig. \ref{fig:soft-label asr} shows a consistent decline of ASR over the increase in noise level $\sigma$. This steady decline indicates robust defense performance of the BD method against the soft-label attacks.
%%%%%%%%%%%%%%%%%%%%%%%%%%%%%%%%%%%%%%%%%%%%%%%%%%%%%%%%%%%%%%%%%%
\subsubsection{ASR of hard-label black-box attacks}\label{hard-label section}

%To evaluate the performance of BD method against hard-label-based attacks, we considered same threshold ($\theta$) and noise ($\sigma$) that was used for the soft-label attacks, for the better understanding and uniformity in the performance evaluation. 

We have summarized the ASR and median $l_2$ distortion of hard-label attacks in presence of our proposed BD method in Table \ref{tb2:ASR2}. 

\begin{table}[t]
	\caption{ASR (\%) and Median $l_2$ Distortion of Targeted Hard-label Attacks. $\sigma =0.1$. ``-" Means no $l_2$ Distortion Data due to Absence of Adversarial Samples.}
	\label{tb2:ASR2}
	\centering
	\begin{tabular}{c|c|ccc}
		\toprule
		Dataset     & Attacks & & ASR/$l_2$& \\
		&  & No defense & $\theta_1$ = 0.5 & $\theta_2$ = 0.7\\
		\midrule
		       & Sign-OPT &  100/0.059  & 4/0.12 & 0/-\\
		MNIST  & BA & 100/0.16 & 17/0.55 & 9/0.56\\
		       & HSJA & 100/0.15 & 38/0.14 & 7/0.15\\
		 \midrule
		  CIFAR10    
		      & Sign-OPT & 100/0.004 & 4/0.08  &  0/-\\
		      & HSJA & 100/0.05 & 18/0.05 & 7/0.05\\
		 \midrule \midrule
		 Dataset     & Attacks   & &ASR/$l_2$&\\
		     &   & No Defense & $\theta_1$ = 0.1 & $\theta_2$ = 0.3\\
		    \midrule
		  
		      & NES-HL & 90/0.12  & 0/- &  0/- \\
	IMAGE-   & Sign-OPT & 100/0.05  & 14/0.4 & 0/- \\
	NET	      & BA & 100/0.08  & 0/- & 0/- \\
		      & HSJA & 100/0.03  & 34/0.11 & 0/- \\
		\bottomrule
	\end{tabular}
\end{table}

Surprisingly, the BD method performed extremely well against the hard-label attacks that were usually challenging to conventional defense methods.  In general, BD(0.3, 0.1) was able to reduce ASR to 0\% over the IMAGENET dataset, and BD(0.7, 0.1) was enough to reduce ASR to near 0 over MNIST and CIFAR10. 

\begin{figure}[t]
	\centering
    \includegraphics[width=1\linewidth]{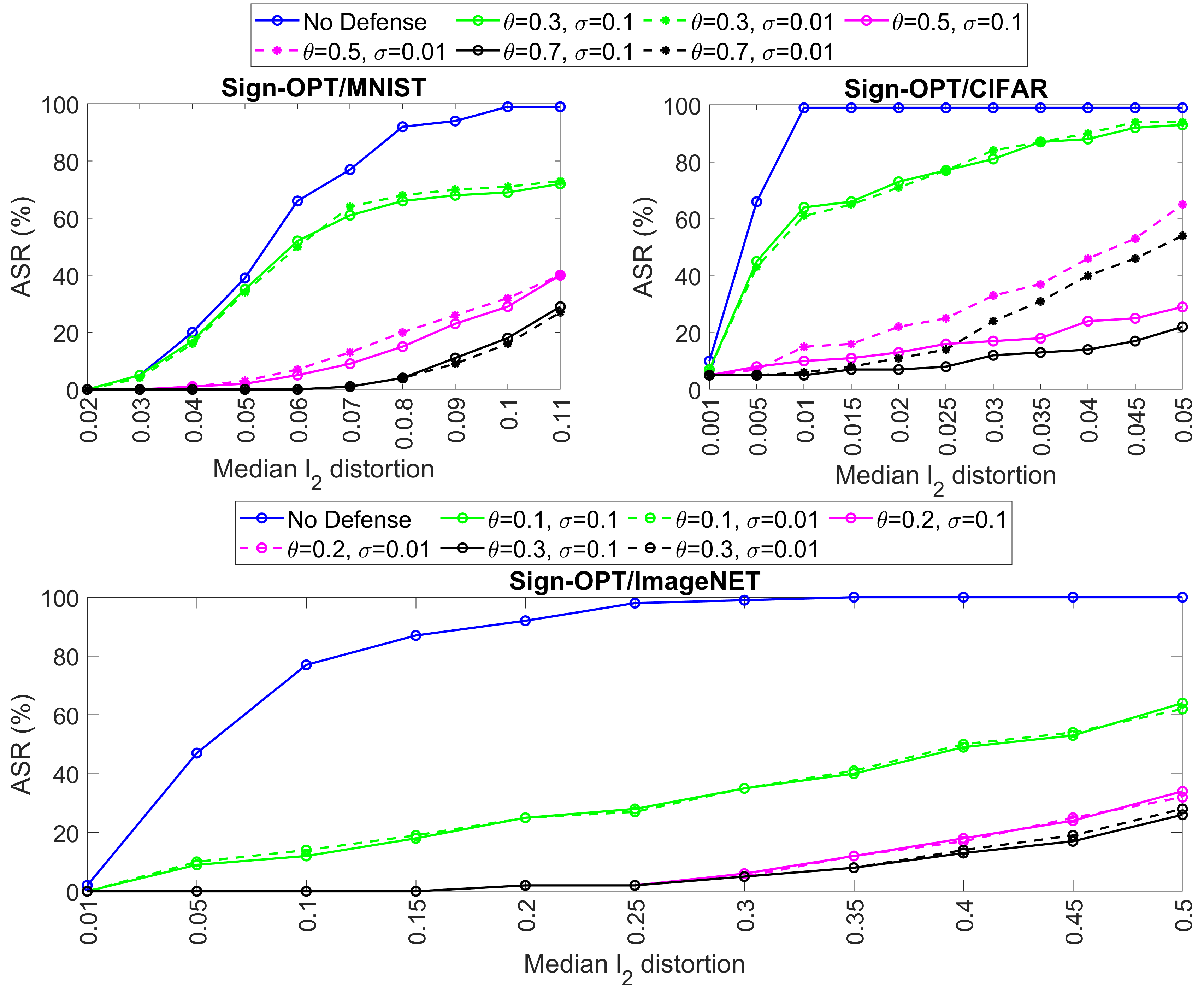}
	\caption{ASR (\%) versus median $l_2$ distortion of the Sign-OPT attack under the proposed BD method. }
	\label{fig:hard-label asr}
\end{figure}

%We have observed that for hard-label methods, the $l_2$ distortion vs ASR analysis was adopted to evaluate the attack effectiveness. Thus, following an inclusive approach was critical for the better analysis of BD method. 
For ASR2, Figure \ref{fig:hard-label asr} shows how ASR2 varies with the pre-set $l_2$ distortion threshold when the BD method was used to defend against the Sign-OPT attack. We can see that the ASR2 reduced with the increase of either $\theta$ or $\sigma$, or both. BD(0.7, 0.1) and BD(0.3, 0.1) successfully defended against the Sign-OPT attack over the MNIST/CIFAR10 and IMAGENET datasets, respectively. %This corroborates our hypothesis and also the experimental results.
%% Increase image font size

%%%%%%%%%%%%%%%%%%%%%%%%%%%%%%%%%%%%%%%%%%%%%%%%%%%%%%%%%%%%%%%%%%
\subsubsection{Robust defense performance against adaptive attacks} \label{adaptive_attack}
To evaluate the robustness of the defense, it is crucial to evaluate the defense performance against adaptive attacks \cite{tramer2020adaptive}. For example, the attacker may change the query limit or optimization step size. In this subsection, we show the effectiveness of our BD defense against 2 major adaptive attack techniques: 1) adaptive query count (QC) budget; and 2) adaptive step size. 

\begin{figure}[t]
	\centering
    \includegraphics[width=1\linewidth]{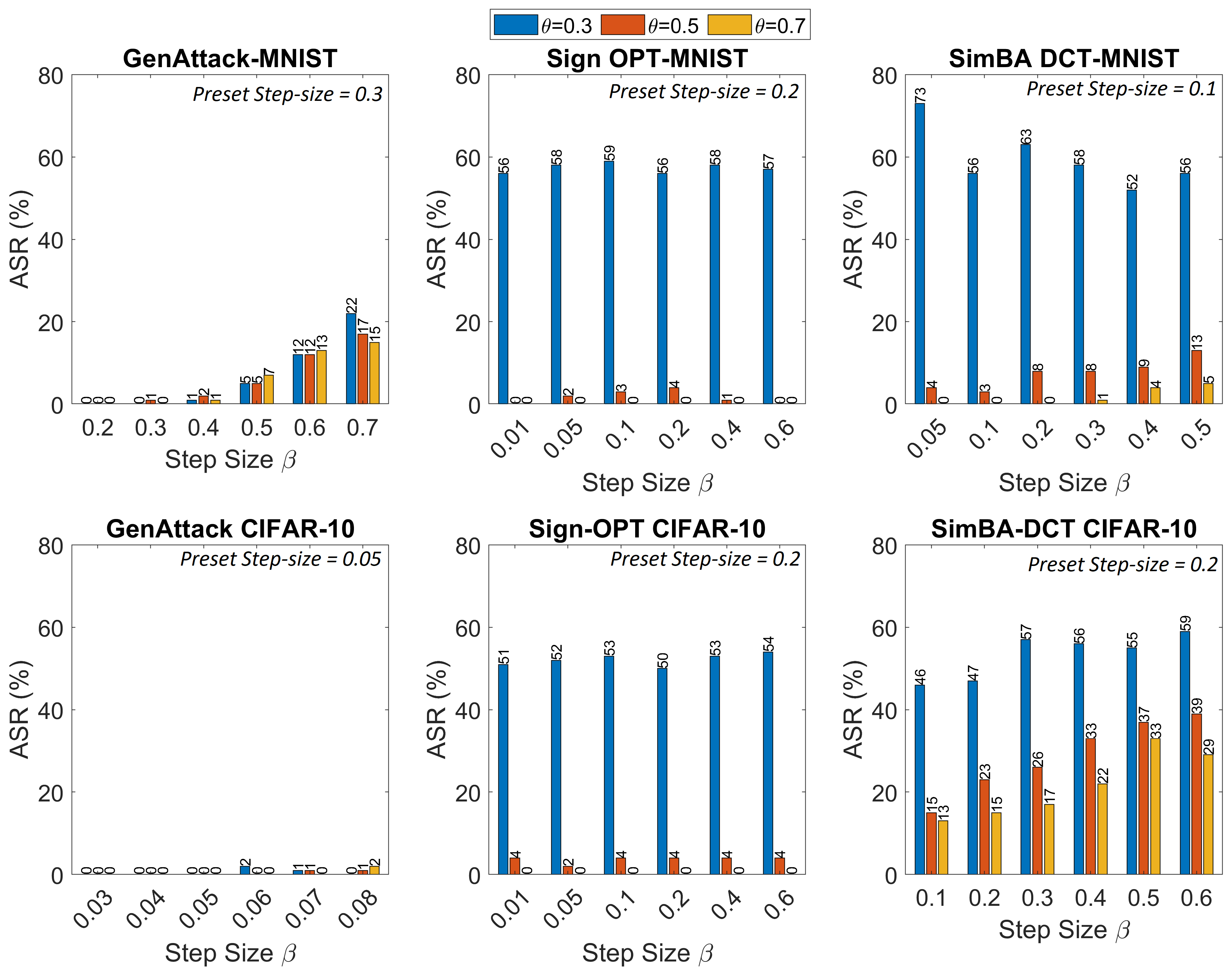}
	\caption{ASR (\%) versus step size of the adaptive attacks. Note that for GenAttack we considered $\sigma=0.01$, since the ASR for $\sigma=0.1$ was always 0 for all threshold values. For Sign-OPT \& SimBA we considered $\sigma=0.1$.}
	\label{fig:step_size}
\end{figure}

First, with increased attack QC budget, the results obtained are summarized in Table \ref{tb3:eot}. %We considered increasing the attacker's query budget much higher than the preset value for the attack methods and ran it along with our best parameter settings (i.e., $\theta$ = 0.5 and $\sigma$ = 0.1 for CIFAR10, and $\theta$ = 0.3 and $\sigma$ = 0.1 for IMAGENET). We adopted query budget as $\{10^4 ({\rm preset}), 10^8, 10^{10}\}$ in this study. 
%Due to nature of the attack methods and larger size of ImageNET data, we reduced the number of test images for this part of experiments to 50 randomly selected images.
We observe that when the attacker increased QC from $10^4$ to $10^{10}$, there was no significant increase in ASR. %In fact, for some cases, like HSJA \& NES-QL, we observed further reduction in the ASR values and for other cases, the ASR values stayed constant even after significant sacrifice of the query budget. 
Next, we adjusted the optimization (or gradient estimation) step size of the attack algorithms (such as $\beta$ of the Sign-OPT algorithm), and evaluated the performance of BD.  The ASR data are shown in Fig \ref{fig:step_size}. We can see that there was no significant change of ASR when the attack algorithms adopted different optimization step sizes. For GenAttack \& Sign-OPT, the ASR was almost the same under various step sizes. For SimBA, the ASR slightly increased but with an expense of heavily distorted adversarial output. 
%To save the space, visual illustration of adversarial outputs with change in step-size is provided in supplementary material.

\begin{table}[t]
	\caption{ASR(\%) of Adaptive Black-Box Attacks under the Proposed BD method}
	\label{tb3:eot}
	\centering
	\begin{tabular}{c|c|ccc}
		\toprule
		Dataset    & Attack &  query & budget &\\
          &   & $10^4$ (preset)  &  $10^8$ & $10^{10}$\\
       \midrule
		 & GA  & 0 & 0 & 0 \\

	     CIFAR10 & HSJA & 3 & 4 & 0 \\
		
		  & Sign-OPT & 5 & 8 & 8  \\
		\midrule
		 & NES-QL & 2 & 12 & 8 \\
		 
		 ImageNET & Boundary & 0  &  0 & 0\\
		 
		  & HSJA & 0 &  0 & 0 \\
		 
		  & Sign-OPT & 0 & 0 & 0\\
		\bottomrule
	\end{tabular}
\end{table}

%We also carefully selected a range of step  size for this part of analysis and the test was conducted for 100 randomly selected images from MNIST \& CIFAR-10 data.

As a result, we can assert the robustness of the BD method against the black-box adversarial attacks.

\begin{table}[h]
	\caption{Comparison of BD method with other defense methods against targeted hard-label attacks in terms of ASR (\%).}
	\label{tb5:other_defense}
	\centering
	\begin{tabular}{c|c|ccc}
		\toprule
		Dataset     & Defense & HSJA & BA & SimBA-DCT \\
		\midrule
		     & DD \cite{papernot2016distillation} & 98 & 80 &  - \\
		     & AT \cite{goodfellow2014explaining} & 100 & 50 &  4 \\
		      & Region-based \cite{cao2017mitigating} & 100 & 85 & - \\
		    MNIST &  BD ($\theta = 0.5, \sigma = 0.1$)
		    & \textbf{38}& \textbf{17}  & \textbf{3} \\
		      & BD ($\theta = 0.7, \sigma = 0.1$)&  \textbf{7} & \textbf{9}& \textbf{0} \\
		\bottomrule
	\end{tabular}
\end{table}
\begin{table}[t]
	\caption{Comparison of BD method ($\theta=0.5, \sigma=0.1$) with other defense methods against targeted GenAttack (soft-label) in terms of ASR (\%).}
	\label{tb6:defenses}
	\centering
	\begin{tabular}{c|c|ccccc}
	
		\toprule
		Dataset     & Attack  & Bit-Depth & JPEG & TVM & NP & BD\\
		\midrule
				 MNIST & GenAttack & 95 & 89 & - & 5 & \textbf{0}\\
		CIFAR10 & GenAttack & 95 & 89 & 73 & 6 & \textbf{0}\\

		\bottomrule
	\end{tabular}
\end{table}
\begin{table}[t]
	\caption{Compare ASR (\%) of proposed BD method with the Robust Bench Defense Models. CIFAR10 Dataset.}
	\label{tb4:robustbench}
	\centering
	\begin{tabular}{c|ccc}
		\toprule
     RobustBench Defense  & Sign-OPT & SimBA & HSJA \\
		\midrule
		      RMC \cite{wu2020adversarial} & 100 & 83 &  100 \\
		       RATIO \cite{augustin2020adversarial} & 100 & 59 & 100 \\
		       RO \cite{rice2020overfitting} & 100 & 85 & 100 \\
		       MMA \cite{ding2018mma} & 100 & 83 & 100 \\
		       ER \cite{engstrom2019adversarial} & 100 & 92 &  100\\
 RD \cite{rony2019decoupling} & - & 80  & - \\
		       PD  \cite{sehwag2021improving}& 100 & 71 &  100\\
		       BD ($\theta = 0.5, \sigma = 0.1$) & \textbf{4}  & \textbf{23} & \textbf{18}\\
		       BD ($\theta = 0.7, \sigma = 0.1$) & \textbf{0} & \textbf{15}& \textbf{7}\\
		\bottomrule
	\end{tabular}
\end{table}

%%%%%%%%%%%%%%%%%%%%%%%%%%%%%%%%%%%%%%%%%%%%%%%%%%%%%%%%%%%%%%%%%%

\subsection{Comparison with Other Defense Methods}

For defending against hard-label attacks, Table \ref{tb5:other_defense} compares the BD method with the DD, AT, and Region-based defense methods over the MNIST dataset. We obtained these other methods' defense ASR data from \cite{chen2020hopskipjumpattack} for HopSkipJump and BA attack methods, and obtained the defense performance data against SimBA through our experiments. It can be seen that our BD method outperformed all these defense methods with lower ASR. For soft-label attacks, Table \ref{tb6:defenses} shows that our BD method also outperformed a list of existing defense methods. 

We also ran experiments using RobustBench models.  The defense performance over the CIFAR10 dataset is reported in Table \ref{tb4:robustbench}.  ASR is used as our preliminary evaluation criteria because for an attacker the higher the ASR the more robust the attack method is against all the defenses. From Table \ref{tb4:robustbench}, we can see that our method had the most superior defense performance.

%%%%%%%%%%%%%%%%%%%%%%%%%%%%%%%%%%%%%%%%%%%%%%%%%%%%%%%%%%%%%%%%%%%%%%%%%%%%%%%%%%%%%%%%%%%%%%%%%%%%%%%%%%%%%%%%

\section{Conclusions} \label{conclusion}

In this paper, we propose an efficient and effective boundary defense method BD$(\theta, \sigma)$ to defend against black-box attacks. This method detects boundary samples by examining classification confidence scores and adds random noise to the query results of these boundary samples. BD$(0.3, 0.1)$ is shown to reduce the attack success rate to almost 0 with only about 1\% classification accuracy degradation for IMAGENET models. Analysis and experiments were conducted to demonstrate that this simple and practical defense method could effectively defend the DNN models against state-of-the-art black-box attacks.

%%%%%%%%%%%%%%%%%%%
\bibliographystyle{IEEEtran} %_fullname}
\bibliography{IEEEabrv,my.bib}

\newpage $\ $
\newpage
%%%%%%%%%%%%%%%%%%%%%%%%%%%%%%%%%%%%%%%%%%%%%%%%%%%%%%%%%%%%%%%%%%%%%%%%%%%%%%%%%%%%%%%%%%%%%%%%%%%%%%%%%%%%%%%%
\centerline{\large {\bf Supplementary Material}}

\appendix

\subsection{Supplementary Experiment Results for Defense Against Targeted Attacks}
\subsubsection{Median $l_2$ Distortion of Soft-Label Targeted Attacks}
In Table \ref{tbl:l21}, we have listed the median $l_2$ distortion of targeted soft-label attacks when our proposed BD$(\theta, \sigma)$ defense was applied. Note that when the ASR was 0\%, then there were no successful adversarial samples generated, and thus the $l_2$ distortion was absent. From Table \ref{tbl:l21}, we can see that there is no significant change in the median $l_2$ distortion when comparing the cases with defense and without defense. This is the same as the result of hard-label attacks shown in Table \ref{tb2:ASR2}. %For some attacks, we observed slight decrease in distortion value

\begin{table}[h]
	\caption{Median $l_2$ Distortion of Targeted Soft-Label Attacks under the BD$(\theta, \sigma)$ defense. $\sigma = 0.1$}
	\label{tbl:l21}
	\centering
	\begin{tabular}{c|c|ccc}
	
		\toprule
		 Dataset   & Attacks  & No defense & $\theta_1$ = 0.5 & $\theta_2$ = 0.7 \\
		  	\midrule
	 & AZ & 0.0473 & 0.0478 & 0.0478  \\
     MNIST   & GenAttack & 4.8086 & - & -\\
     & SimBA & 0.0957 & 0.0694 &  0.0823 \\
		 \midrule
	  &  AZ  & 0.0672 & 0.0672  &  0.0672 \\
	 CIFAR10    & GenAttack & 1.3874 & - & -  \\
	  & SimBA & 0.0323 & 0.0236 & 0.0218\\
	   \midrule \midrule
		 Dataset     & Attacks & No Defense & $\theta_1$ = 0.1 & $\theta_2$ = 0.3 \\
		  \midrule
	  &  AZ  & 0.1566 & - &  - \\
	IMAGENET   & GenAttack & 0.0377 & - & -  \\
	  & SimBA & 0.0163 & 0.0092 & 0.0075\\

		\bottomrule
	\end{tabular}
\end{table}

\subsubsection{Visual Representation of Targeted Attack Samples}
\begin{figure}[h]
	\centering
    \includegraphics[width=1\linewidth]{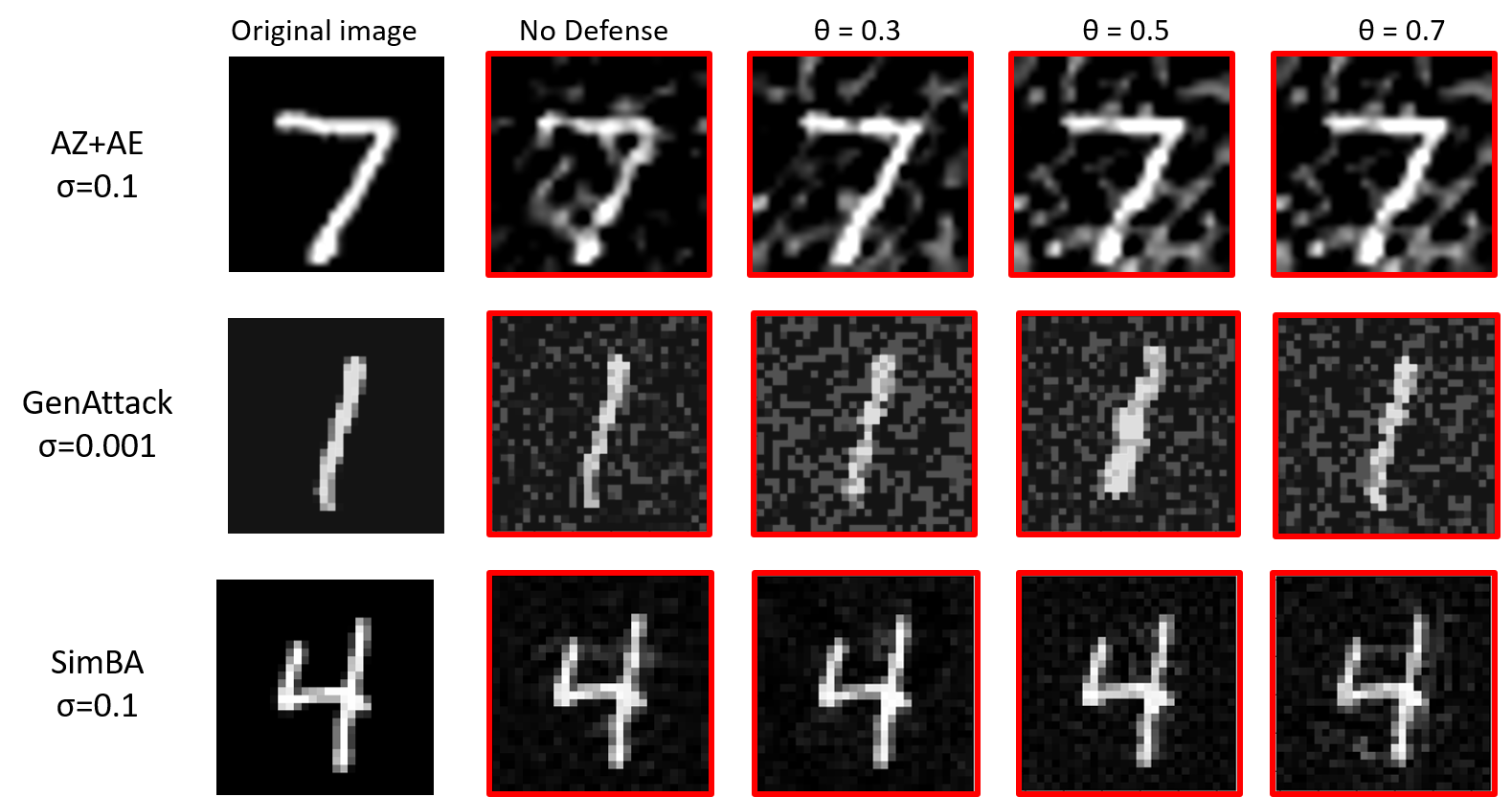}
	\caption{Adversarial samples generated by the targeted soft-label attacks under the BD($\theta,\sigma$) defense over the MNIST dataset.}
	\label{fig:visualrep_1}
\end{figure}

\begin{figure}[h]
	\centering
    \includegraphics[width=1\linewidth]{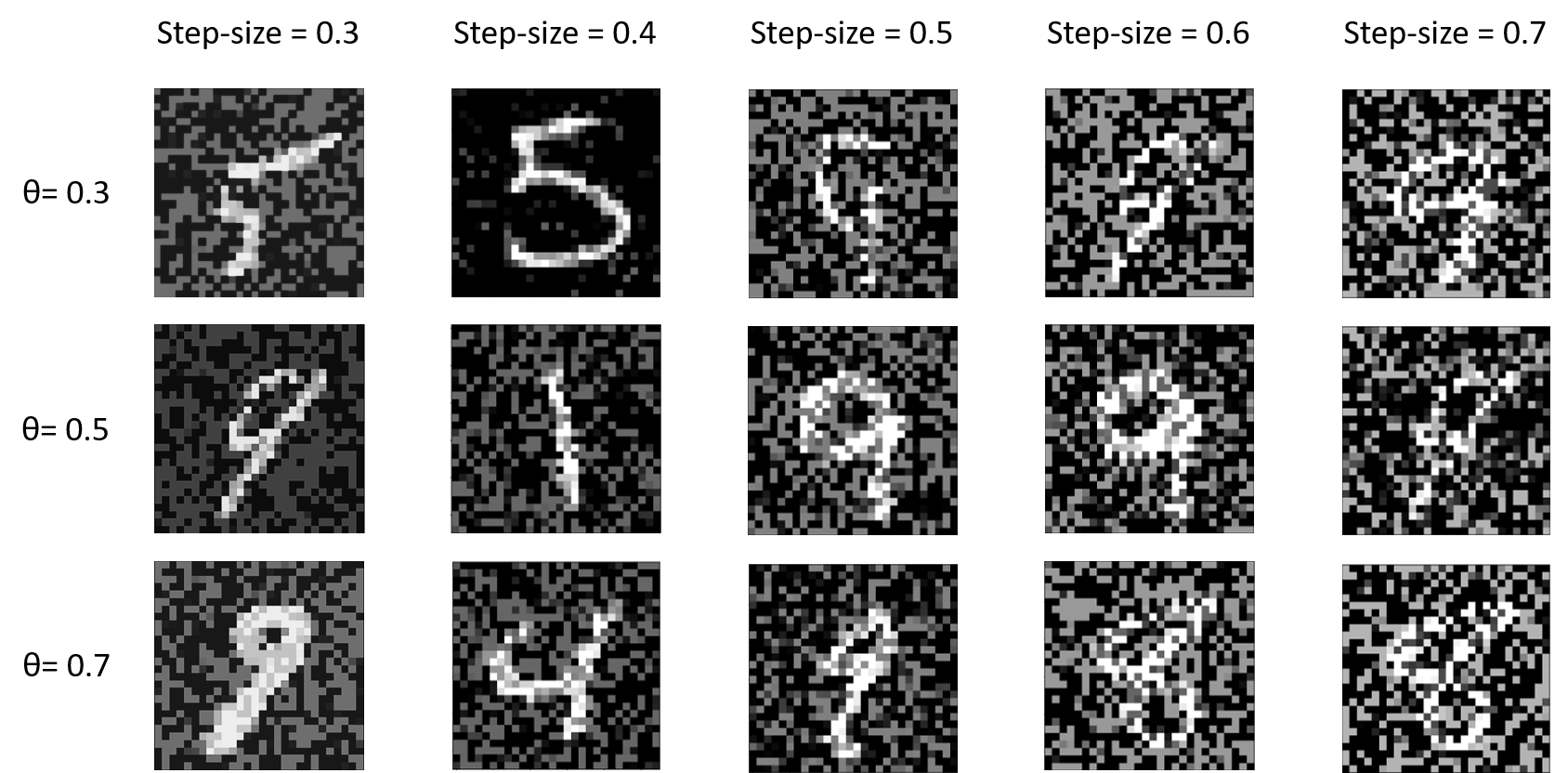}
	\caption{Adversarial samples generated by the GenAttack with adaptive attack techniques (various optimization step-sizes) under the BD$(\theta, 0.1)$ defense. Note that the default step-size used in the GenAttack was 0.3 for the MNIST dataset.}
	\label{fig:visualrep_2}
\end{figure}

For the adversarial samples generated by the targeted attack algorithms under our proposed defense, we examine whether the distortion, i.e., the difference between the original sample and the adversarial sample, is imperceptible to the human eye. Some of the adversarial samples are shown in Fig. \ref{fig:visualrep_1} and \ref{fig:visualrep_2}. Fig. \ref{fig:visualrep_1} shows the MNIST adversarial samples generated by the AZ, GenAttack, and SimBA attack methods with our BD$(\theta, \sigma)$ defense. We can observe that when the proposed BD method was used, the adversarial images generated were extremely distorted. 

% show how BD work against BA & HSJA attack by not changing the labels visually

In Section \ref{adaptive_attack}, we showed the robust performance of the proposed BD method against adaptive attacks, where we found that ASR stayed the same or just increased slightly even under the adaptive attacks. For adaptive attacks that changes the optimization step-size to increase ASR, some of the adversarial samples in the presence of the BD method are illustrated in Fig \ref{fig:visualrep_2}. We can see that the cost of slight increase  of ASR was the heavy image distortion that can be detected by human perception easily.

% false sense of success

\subsection{Defense Performance against Untargeted Attacks}
We evaluated the performance of the proposed BD method against untargeted attacks with 1000 validation images of MNIST and 100 validation images of IMAGENET. We considered the following untargeted black-box attacks: soft-label attacks - \textbf{AZ} \cite{tu2019autozoom}, \textbf{SA}(Square Attack) \cite{andriushchenko2020square}, and \textbf{SimBA} \cite{guo2019simple};  hard-label attacks - \textbf{BA}(Boundary Attack) \cite{brendel2017decision}, \textbf{Sign-OPT} \cite{cheng2019sign}, and \textbf{HSJA}(HopSkipJump) \cite{chen2020hopskipjumpattack}. Similar to the targeted case, we adopted their original source code with the default hyper-parameters and considered the $l_2$ norm setting throughout the experiment. 

Besides evaluating the defense performance of our BD method against the above mentioned untargeted attacks, we have also compared the BD method's performance with other existing defense methods, including \textbf{RND} (Random Noise Defense) \cite{qin2021random},\textbf{GF} (Gaussian Augmentation Fine-Tuning) \cite{qin2021random}, \textbf{FD} (Feature Denoising) \cite{qin2021random}, and \textbf{AT} (Adversarial Training) \cite{goodfellow2014explaining}. 
%whose data were obtained from \cite{qin2021random} for the IMAGENET dataset.

An important aspect to consider for the untargeted attack case is the performance metric ASR. For the untargeted attacks, the attacker can simply add enough noise to the original sample to make the adversarial sample satisfy $\arg \max_i F_i({\bf X}) \neq c$. Since all the defense methods in general make the attack algorithms to generate samples with heavy distortion or noise, the measure $\arg \max_i F_i({\bf X}) \neq c$ tends to be satisfied by all the adversarial samples trivially. As a result, 
the first ASR definitions we used in the targeted attack case earlier in Section \ref{experiment} is not very meaningful. In this subsection, we use the second ASR definition, i.e., ASR2, which was defined in Section \ref{experiment} as the ratio of samples satisfying both $\arg \max_i F_i({\bf X}) \neq c$ and median $l_2$ distortion less than a certain distortion threshold $L$. %Compared with ASR2, we drop the $\arg \max_i F_i({\bf X}) \neq c$ measure. 

To maintain a fair calculation of ASR with respect to the $l_2$ distortion threshold $L$ for all the attack algorithms, we determined the threshold value $L$ for all the attack methods from their no-defense results. Specifically, with their original source code and default parameters, we determined a minimum threshold that makes ASR to be near 100\%. Then, we selected the $L$ value to be multiple folds of this minimum threshold value. For example, for the AZ method we used distortion threshold $L = 0.5$ to calculate ASR for MNIST and $L= 1.5$ for IMAGENET. These $L$ values were $10 \times$ larger than the minimum threshold. The main objective was to guarantee almost 100\% ASR for all the attack methods without defense. We used the same $L$ value to calculate the ASR when the BD method was applied.

\subsubsection{ASR of Soft-Label Untargeted Attacks}

The ASR of the soft-label untargeted black-box attacks under our proposed BD method is summarized in Table \ref{tbl:ASR3}. We have shown the experiment data for $\sigma=0.1$ and varying $\theta$ values. We can see that when defending against untargeted soft-label attacks, the performance of the BD method indicated the same trend as what we observed for targeted attacks. The ASR degraded drastically with the increase of $\theta$. BD$(0.7, 0.1)$ defended the MNIST DNN models against untargeted attacks as effectively as against targeted attacks. However, for the IMAGENET DNN models, as compared to the targeted attack case, we had to increase the threshold $\theta$. This is not surprising considering the fact that, compared with the targeted attack, the untargeted attack is usually easier for the attacker to be successful and is thus harder for the defender to mitigate. BD$(0.4, 0.1)$ reduced the ASR of AZ and SimBA to 0, and reduced the ASR of SA to 30\%. Increasing $\theta$ further to 0.6 would reduce the ASR of SA to 15\% as shown in Table \ref{tb9:other_defense_2}. Note that such a moderate increase of $\theta$ would not lead to too large ACC degradation, as we have evaluated in Section \ref{accevaluation}.

\begin{table}[t]
	\caption{ASR (\%) of Untargeted Soft-Label Attacks. $\sigma$ = 0.1.}
	\label{tbl:ASR3}
	\centering
	\begin{tabular}{c|c|ccc}
	
		\toprule
		Dataset     & Attacks   %&  & ASR (\%) &\\ & 
		& No defense & $\theta_1$ = 0.5 & $\theta_2$ = 0.7 \\
		      % \cmidrule{3-5}
		     % &  &  &  & \\
		\midrule
		 & AZ &  97 & 0 & 0  \\
		  MNIST    & SA & 92 & 89 & 10\\
		       & SimBA & 82 & 31 & 0 \\
		 \midrule \midrule
		 Dataset     & Attacks   %&  &ASR (\%) &\\
		     %&   
		     & No Defense & $\theta_1$ = 0.3 & $\theta_2$ = 0.4 \\
		  \midrule
		  & AZ & 100 & 0 & 0  \\
		  IMAGENET & SA & 100 & 38 & 30\\
		      & SimBA &  100 & 96 & 0 \\
		\bottomrule
	\end{tabular}
\end{table}

\subsubsection{ASR of Hard-Label Untargeted Attacks}

\begin{figure*}[t]  %% add theta = 0.3 in graph asap
	\centering
    \includegraphics[width=0.9\linewidth]{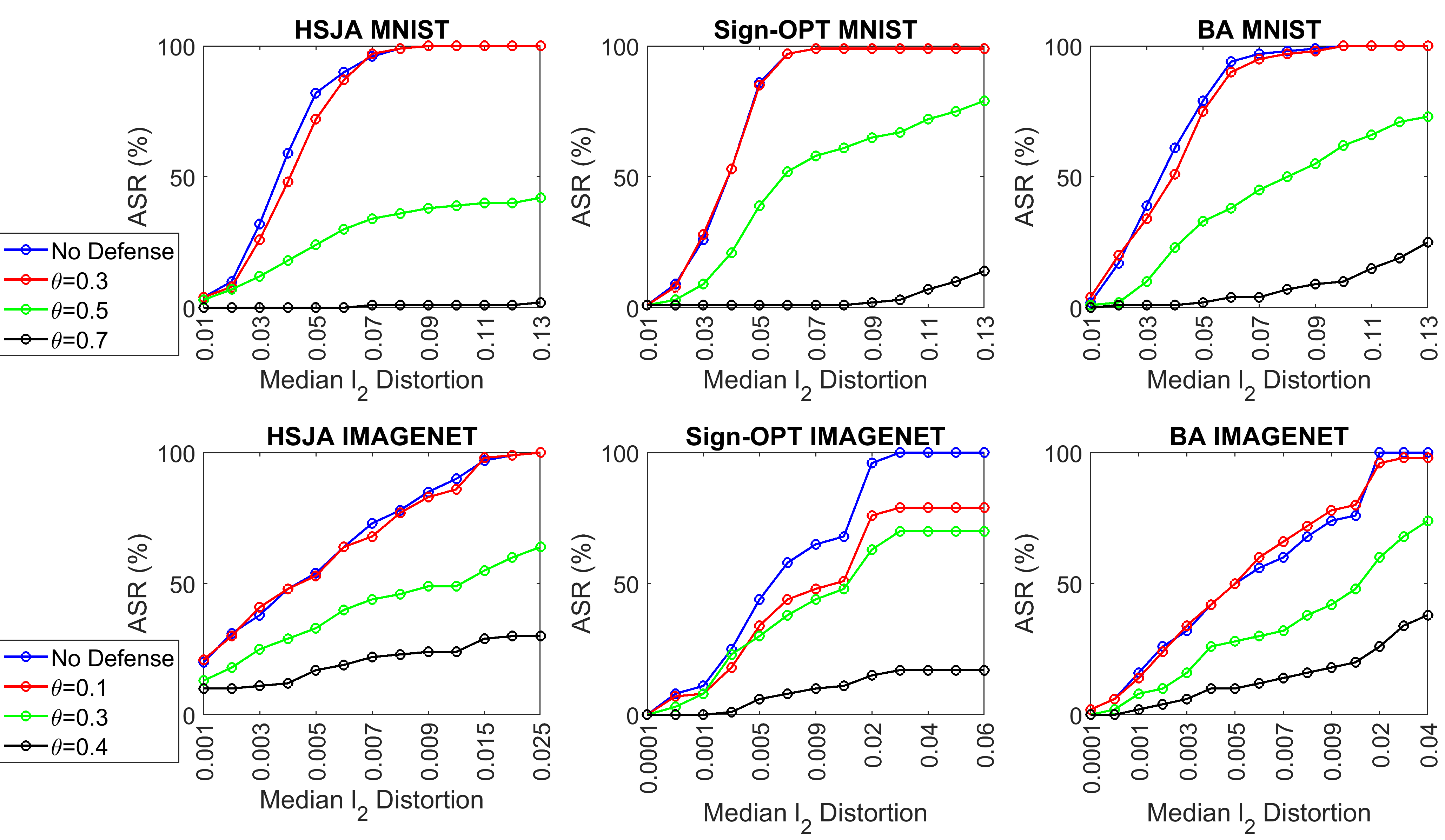}
	\caption{ASR (\%) vs. median $l_2$ distortion threshold $L$ of untargeted hard-label attacks under the proposed BD method BD$(\theta, 0.1)$.}
	\label{fig:hard-label-UT}
\end{figure*}

We have summarized the ASR of hard-label untargeted attacks in the presence of the proposed BD method in Fig. \ref{fig:hard-label-UT}. As observed in the targeted attack case, the ASR of the untargeted attack methods display a steady decline with the increase of $\theta$. %respect to the $l_2$ distortion threshold  with the increase in $\theta$ value.

\subsubsection{Comparison with Other Defense Methods against Untargeted Attacks}

\begin{table}[t]
	\caption{Comparison of the BD method with other defense methods against untargeted black-box attacks in terms of ACC degradation and ASR} %The defense performing best while maintaining the lowest model ACC degradation is highlighted in bold.}
	\label{tb9:other_defense_2}
	\centering
	\begin{tabular}{c|c|c|cc}
		\toprule
		Dataset     & Defense & Degradation & SA & SimBA\\
		    & Method & of ACC (\%) & ASR (\%) & ASR (\%)\\
		\midrule
		     & RND \cite{qin2021random} & -1.9 & 51.9 & 38.8  
		     \\
		     & GF \cite{qin2021random} & -0.2 & 95.8 & 88.8 \\
		   IMAGE-   & FD \cite{qin2021random} & -20.8 & 51.8 & 38.7 \\
		    NET  & AT \cite{goodfellow2014explaining} & -13.4 & 47.2 & 35 \\
		    & RND+AT \cite{qin2021random} & -16.8 & 20 
		    & 4.7 \\
%		     &  BD ($\theta = 0.4,$ & &  &   \\
		    &  BD(0.4, 0.1) & -2.0 &\textbf{30}  & \textbf{0}\\
%		     &  BD ($\theta = 0.6,$ & &  &   \\
		    &  BD(0.6, 0.1) & -3.5 &\textbf{15}  & \textbf{-}\\
		\bottomrule
	\end{tabular}
\end{table}

We have compared the performance of the BD method with some existing defense methods for defending against untargeted black-box attacks. The performance comparison data are illustrated in Table \ref{tb9:other_defense_2}, where most of the data of the existing defense methods were obtained from \cite{qin2021random}. We can see that  with 2 to 3.5\% ACC degradation, the BD method could reduce ASR to near 0. 
%The clean model classification accuracy in \cite{qin2021random} was 74.9\% for ResNet-50 model but the clean accuracy of ResNet-50 model we obtained in our study is 80\%. Please note that we have used Inception-v3 model in our experiments and the clean model ACC we obtained is 79\%. 
The ASR of the Square Attack was comparatively higher and all other defence methods were not effective except RND+AT and our BD method. However, RND+AT suffered from a heavy ACC degradation of -16.8\%, which means this defense method is not useful in practice. Our BD method could suppress the ASR to 15\% using $\theta=0.6$ and $\sigma=0.1$ with only a small -3.5\% ACC degradation.  Thus, we can claim that, in comparison with other defense methods, our BD method demonstrated effective and superior performance to defend against untargeted attacks.

\begin{figure*}[t]
	\centering
    \includegraphics[width=1\linewidth]{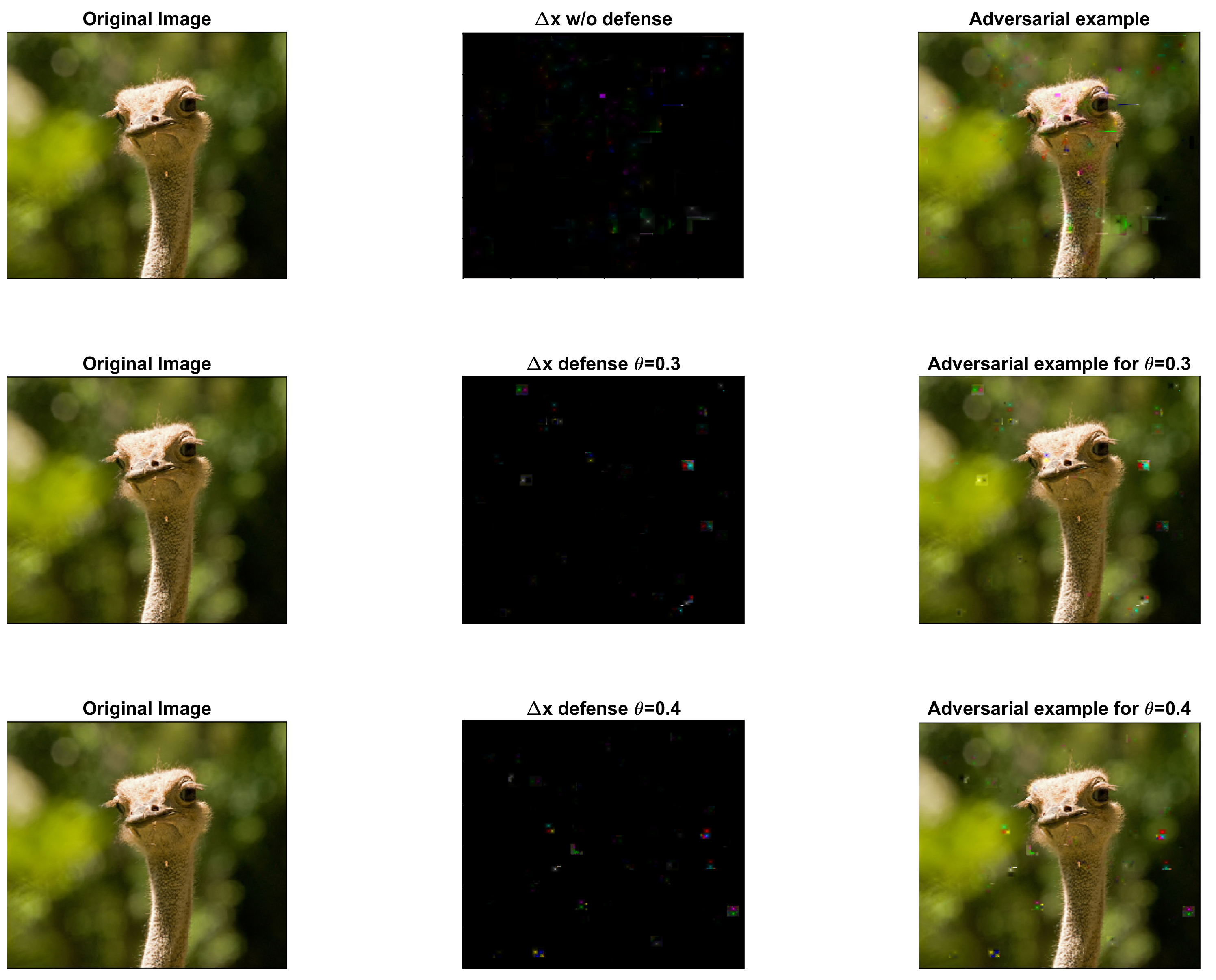}
	\caption{Original images ${\bf X}_0$, distortions $\Delta {\bf x}$ and adversarial images ${\bf X}$ generated by the Square Attack (untargeted soft-label attack) under the proposed defense BD$(\theta, \sigma)$ for $\theta = \{0, 0.3, 0.4\}$, and $\sigma = 0.1$.}
	\label{fig:sq_diff}
\end{figure*}

\subsubsection{Visual Representation of Untargeted Attack Samples}
\begin{figure*}[t]
	\centering
    \includegraphics[width=0.75\linewidth]{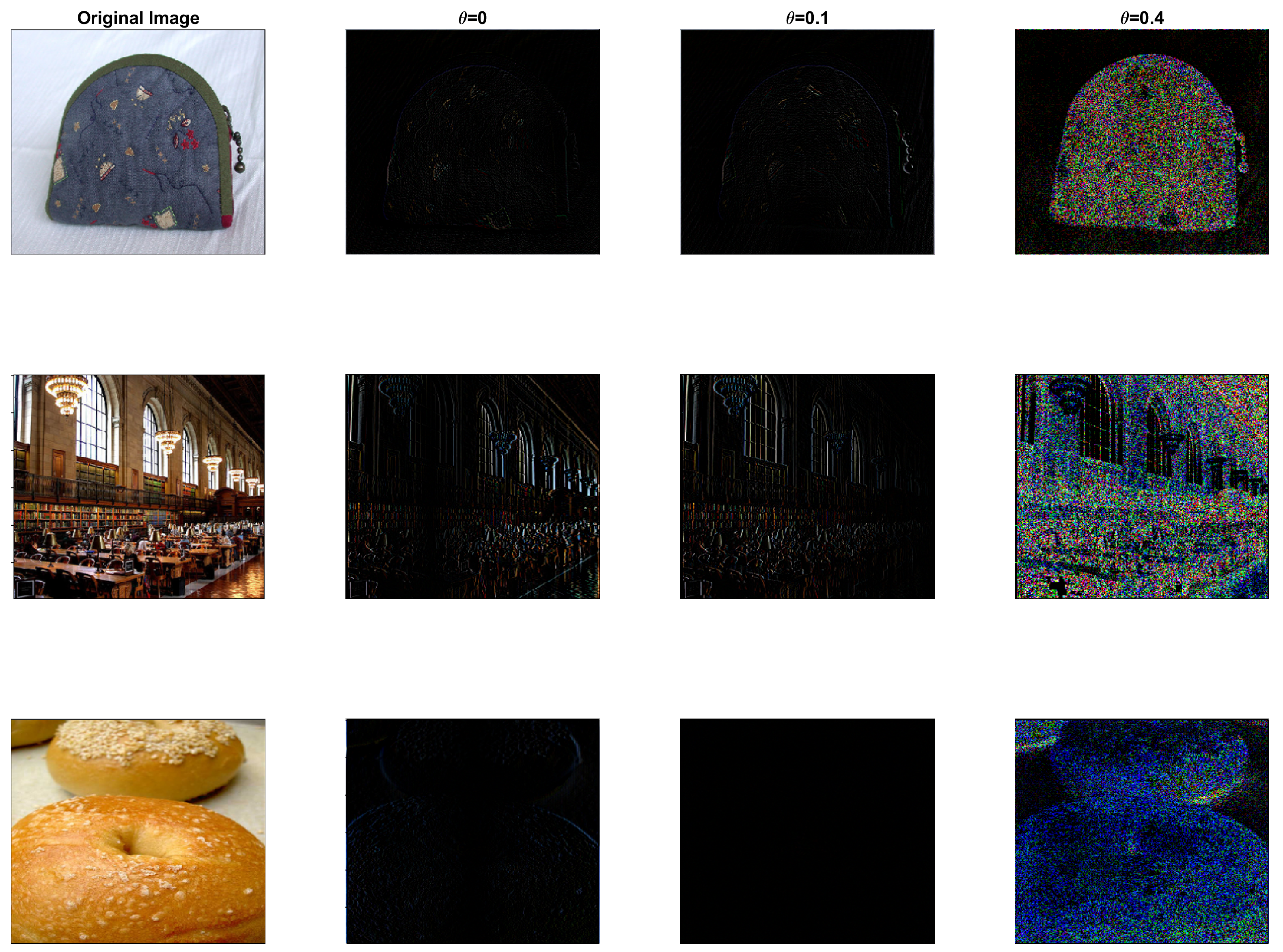}
	\caption{Original images ${\bf X}_0$ and distortions $\Delta {\bf x}$ generated by the Boundary Attack (untargeted hard-label attack) under the proposed defense BD$(\theta, \sigma)$ for $\theta = \{0, 0.1, 0.4\}$, and $\sigma = 0.1$.}
	\label{fig:ba_diff_img}
\end{figure*}

\begin{figure*}[t]
	\centering
    \includegraphics[width=0.75\linewidth]{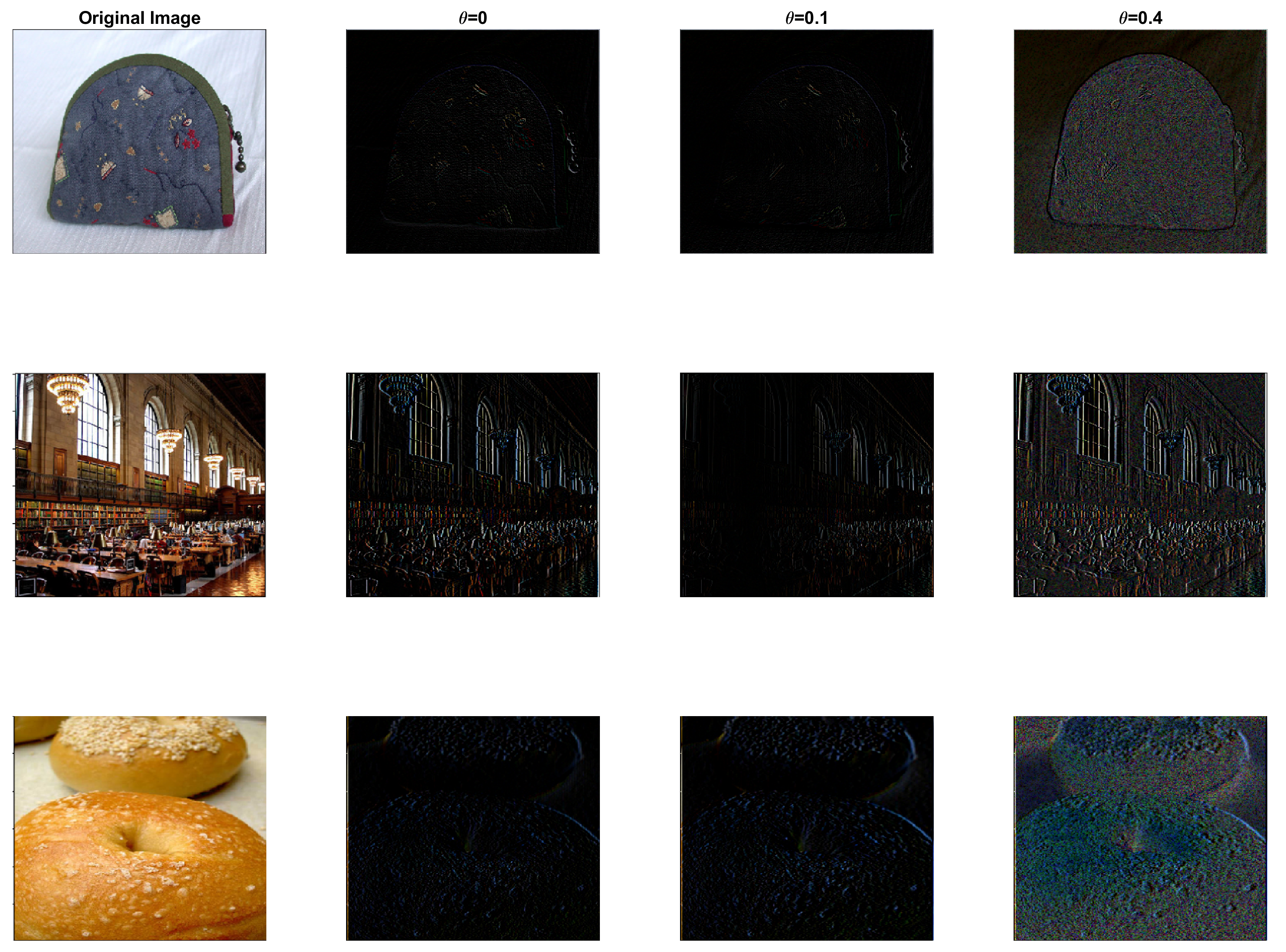}
	\caption{Original images ${\bf X}_0$ and distortions $\Delta {\bf x}$ generated by the HopSkipJump Attack (untargeted hard-label attack) under the proposed defense BD$(\theta, \sigma)$ for $\theta = \{0, 0.1, 0.4\}$, and $\sigma = 0.1$.}
	\label{fig:hsja_diff_img}
\end{figure*}

The visual representation of the adversarial samples generated by the untargeted attack methods in the presence of the BD method is shown in Fig \ref{fig:sq_diff}-\ref{fig:hsja_diff_img}. Fig. \ref{fig:sq_diff} shows the results of the soft-label attack SA, where the left column is the original image ${\bf X}_0$, the right column is the adversarial image ${\bf X}$, and the middle column shows the distortion $\Delta {\bf x} = {\bf X}-{\bf X}_0$.  Note that $\theta=0$ means no defense is applied. 
%for the better understanding of the degradation in the performance of these attack methods during the optimization stage of the soft-label attack method due the presence of the BD method.
In contrast, Fig. \ref{fig:ba_diff_img} and \ref{fig:hsja_diff_img} are for the hard-label attacks, where the first column shows the original images ${\bf X}_0$, and the rest columns show the distortion $\Delta {\bf x}$.

%Fig. \ref{fig:ba_diff_img} and Fig. \ref{fig:hsja_diff_img} represents the difference image w/o BD method against hard-label methods (i.e., Boundary Attack and HopSkipJump Attack) for varying $\theta$ values.

From these figures, we can see that for untargeted attacks, the attacker needed higher distortion in order to generate a successful adversarial image under our BD method. With the increase of $\theta$, the distortion level added by the attacker increased rapidly, especially for hard-label attacks, which resulted in highly degraded adversarial images. %Thus,  we can say that the BD method is equally robust against the untargeted attack methods.

\subsection{Discussions \& Limitations} 
While the experiment results we obtained strongly corroborate our main hypothesis that \textbf{addition of a small random noise to the logits when the query sample reaches close to the decision boundary will successfully defend black-box attacks}, we have also found some interesting observations which will be helpful for the future study of the defense methods. Amongst various soft-label attacks that we experimented, we found that SimBA-DCT and NES-query-limited for CIFAR-10, and IMAGENET data, respectively, demonstrated the strongest performance against most of existing defense methods including our proposed BD method. Similarly, when evaluating the defense performance against hard-label attacks, we observed that the Boundary Attack and the HopSkipJump Attack performed comparatively stronger for datasets with small image sizes like MNIST \& CIFAR-10 in the presence of the BD defense. However, from Table \ref{tb2:ASR2}, we can see that the BD method demonstrated superior performance against the HopSkipJump targeted attack over the IMAGENET dataset. We observed similar trend in the performance of the BD method against untargeted black-box attacks. While the untargeted attack setting is considered to be in the favor of the attacker, defending against untargeted attacks is considered to be tricky. The ASR of the untargeted attacks was comparatively higher than that of the targeted attacks for the same defense parameter setting. But when $\theta$ was increased, without too much ACC degradation, the ASR could still drastically reduce to almost 0\%. Our proposed BD method has demonstrated robust performance against all the listed black-box attacks and we believe that the performance will be consistent for any other or new black-box attack methods. When compared with the existing defense methods, we observed that the BD method presented better and more reliable performance.

It will be very interesting to evaluate the augmented performance of the BD method in combination with some of the existing robust adversarial defenses. We leave this part of the study for future work. Note that our main focus of this paper is the query and search-based black-box attack method where both the model and dataset are unknown to the attacker. We do not cover the transfer-based attack method in this study (due to the inaccessibility of the training set used for training of the surrogate model in this scenario). We leave the study of the defense against transfer-based attack methods for future work.

\end{document}